\documentclass{iopjournal}

\usepackage{graphicx}% Include figure files
\usepackage{amsmath}
\usepackage[numbers,sort&compress]{natbib}

\newcommand{\Nm}{N_m}
\newcommand{\kB}{k_\text{B}}
\newcommand{\kT}{\kB T}

\newcommand{\code}[1]{{\sc #1}}
\newcommand{\DLMESO}{\code{dl\_meso}}
\newcommand{\DLPOLY}{\code{dl\_poly\_5}}
\newcommand{\nDPD}{\mbox{$n$\text{DPD}}}
\newcommand{\nm}{\text{nm}}

\begin{document}

\articletype{Paper} %	 e.g. Paper, Letter, Topical Review...

\title{Highly coarse-grained polarisable water models for mesoscopic simulations}

\author{Michael A. Seaton$^1$\orcid{0000-0002-4708-573X}, Benjamin T. Speake$^1$\orcid{0000-0002-5690-9470} and Ilian T. Todorov$^{1,*}$\orcid{0000-0001-7275-1784}}

\affil{$^1$Scientific Computing, UKRI Science and Technology Facilities Council, STFC Daresbury Laboratory, Sci-Tech Daresbury, Keckwick Lane, Warrington WA4 4AD, United Kingdom}

\affil{$^*$Author to whom any correspondence should be addressed.}

\email{ilian.todorov@stfc.ac.uk}


\begin{abstract}
Modelling micro- and meso-scopic scale thermodynamic and transport properties of soft condensed matter hinges upon its representation.  This is especially relevant for polar solvents such as water, since these require effective representation of their dielectric nature as driven by molecular charge distributions and molecular network structuring.  The dielectric nature of a medium leads to complex phenomena such as local polarisability response and restructuring near interfaces in reaction to changes in local charge distributions.  Inclusion of such phenomena when using larger-than-atomistic techniques such as coarse-grained molecular dynamics (CG-MD) and dissipative particle dynamics (DPD) is still an open question, to which we provide a novel way to consider and justify the necessary and suitable coarse-graining level, enabling us to compare new polar CG models' performance against that of an underlying atomistic model.  We polarise our previous non-polar \nDPD\ water model to prepare it for use in simulations of liquid electrolytes as well as solvated organic membranes and measure its fitness to serve as a dielectric medium by comparing its properties to those of the TIP3P water model, while simultaneously observing changes to properties already represented well by the non-polar model.
\end{abstract}

\section{\label{sec:level1}Introduction}

Mesoscopic modelling methods such as dissipative particle dynamics (DPD)\cite{fs02,ew17} are capable of modelling, \emph{inter alia}, structured liquids with charged groups (e.g. surfactant micelles)\cite{abdr+18}, electrolytes in aqueous solutions\cite{mao2015}, and polyelectrolyte membranes and gels\cite{santo2021,vishnyakov2021}.  Many simulations at these length and time scales assume a constant dielectric permittivity corresponding to a background medium (e.g. water) to simplify determination of charge-based interactions, with at least some success.  This assumption is unlikely to hold, however, for systems where the local dielectric can vary, such as those with interfaces between different phases (e.g. gas-liquid) or fluids with large discontinuous contrasts in relative permittivity (e.g. between oil and water). Even in systems with a single-phase background fluid, local variations in dielectric permittivity can affect the distribution of ionic species and consequently the stability of formed liquid structures\cite{hendrikse2024b}.

\begin{figure}
    \centering
    \includegraphics[width=0.6\columnwidth]{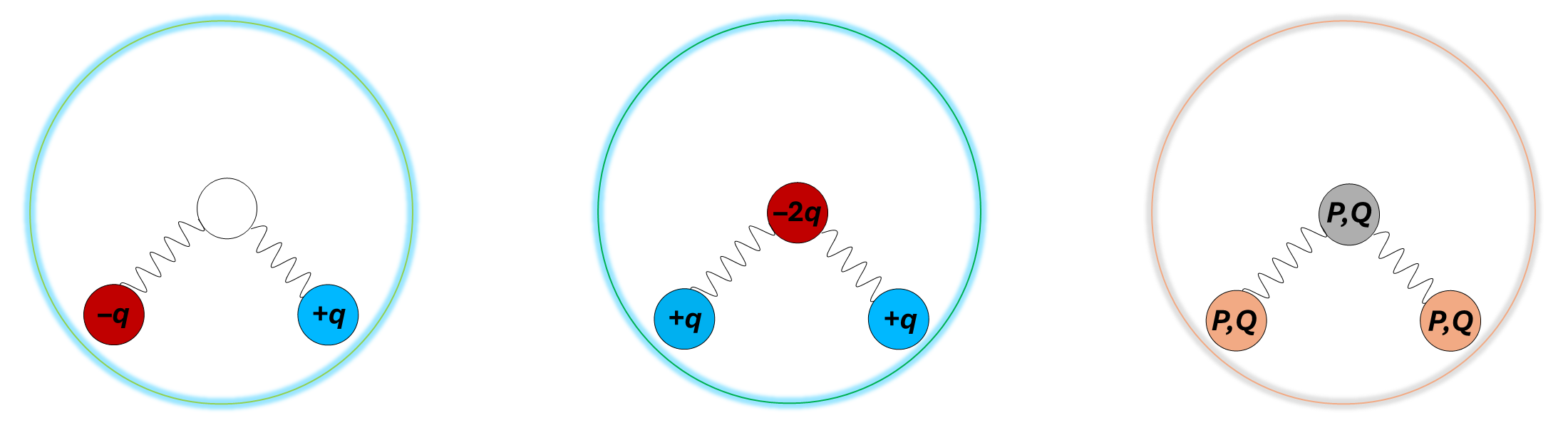}
    \caption{The three general types of 3-site polarising techniques for coarse-grained water models with a central bead site and two satellite sites.  \emph{Left} -- a two-charge model (2CM) with a test charge $q$.  \emph{Middle} -- a three-charge model (3CM), enhancing the dipole with respect to the 2CM by a factor of 2.  \emph{Right} -- a 3-site model, enhanced with constant, point dipole and quadrupole $\{P_s, Q_s\}$ distributions per site (3DM).  Such models usually include harmonic pseudo-bonds (force constant, $\kappa$), with or without equilibrium distances ($r_0$), and may also have a harmonic angular restraint (force constant, $\kappa_\theta$, and equilibrium angle, $\theta_0$).  The models' parameters, including cross-bead interactions, are fitted against a choice of test charge or multipolar distribution to match best the desired permittivity of the medium.}
    \label{3sm}
\end{figure}

To tackle systems with local dielectric variations, the simplest approach available would be to devise models that explicitly incorporate dipole moments for background fluids\cite{pp14,plp15,vjt18,chiacchiera2024,wu2010,li2020} by using three popular techniques summarised in Figure~\ref{3sm}. Starting with charge-neutral particles that interact with short-range pairwise van der Waals-like potentials, additional pseudo-particles can be attached to these using pseudo-bonds to form small molecule-like entities.  Charges are then assigned to the particles in each `molecule' in order to produce a desired dipole moment.  This approach, adding Drude sites to the original beads centres, is significantly simpler than determining local permittivities from solute concentrations\cite{gro03,gro03b} or implicitly imposing permanent dipole moments\cite{li2020}, not least because it uses standard calculation techniques for all charge-based interactions (e.g. Ewald summation) as found in most molecular dynamics (MD) or DPD simulation codes without requiring substantial modifications.  Obtaining the correct dielectric for our background medium of water is evidently a matter of selecting charge magnitudes and bond parameters, although we also seek to obtain realistic charge-based properties such as dipole and quadrupole moment distributions for larger clusters of water molecules, as these play a crucial role near interfaces  where water molecules (and more generally polar liquids molecules) become structurally very highly ordered and drive changes to properties such as surface tension and the structure of the interface as a whole\cite{Croxton1981,abaskal_2007,gongadze2013,cendagorta_2015,Slavchov2015}, as well as the dynamics of dissolved ions\cite{Jeon2003,Chitanvis1996,Slavchov2014,Slavchov2014a}. 

The choice of short-range pairwise (van der Waals-like) interactions between particles representing a solvent at larger-than-atomistic scales needs careful consideration, whether or not dipoles are included.  Some coarse-graining (CG) schemes such as the Martini force field for biomolecular systems\cite{yss+10,souza2021} are based upon Lennard-Jones (12-6 Mie) potential functions, such as those typically used for atomistic simulations. We are not convinced this is a wholly suitable choice for CG simulations, not least since the Lennard-Jones form provides an overly hard repulsion for molecular interactions in polar environments. Even at the atomistic level, the AMOEBA polar force field\cite{ponder2010} adopts a softer 14-7 Mie potential form, which better fits gas phase \emph{ab initio} results and improves the description of hydrogen bonding.

The common choice of softer, entirely repulsive and bounded interaction potentials and forces, as proposed by Groot and Warren\cite{gw97} for mesoscopic simulations based upon DPD, provides a more realistic representation of interactions between beads for the intended length scales and is broadly similar to those obtained by systematic CG from atomistic models\cite{forrest1995}.  This is why it is still widely used for DPD calculations with several parameterisation schemes available\cite{gw97,vln13,abf+17,sepehr2016} to match up interactions with various properties, approaching the mesoscopic scale from either smaller (bottom-up) or larger (top-down) scales.  The simplicity of this `standard DPD' interaction, particularly its lack of attractive terms, unfortunately leads to less realistic thermodynamic behaviour with a quadratic equation of state in terms of particle density and thus no available phase coexistence for a single species.

Extending these interactions to include dependencies on local particle densities, an approach known as `many-body DPD'\cite{pagonabarraga2001,war03}, can improve their connection to real thermodynamics: for instance, these can be selected directly to produce a known equation of state.  The main drawback to this approach is the need for additional pair-based calculations of localised particle densities, used to avoid possible thermodynamic inconsistencies of global density-dependent pair potentials\cite{louis2002}.  Many of the required behaviours can alternatively be obtained using standard pairwise interactions.  One such approach, which we denote as \nDPD\cite{sokhan2023}, modifies the `standard DPD' function to include additional attraction and control on repulsions, leading to realistic vapour-liquid coexistence below a critical point and intriguing solid-liquid transition behaviour, e.g. negative thermal expansion around the melting point.

It is worth noting that in traditional CG models of liquids as the mapping number increases, the liquid phase representation faces significant challenges, including numerical freezing and instabilities.  Specifically, in Martini\cite{marrink2007} and DPD water\cite{trofimov2003PhD} models, crystallisation can occur at comparatively low CG mappings, regardless of whether or not the correct physical viscosity and compressibility for the liquid phase are enforced simultaneously\cite{pk06,trofimov2003PhD}. This freezing is linked to the Kirkwood–Alder transition\cite{DZWINEL00}, observed as the mapping number and speed of sound increase.  The general root cause for this is entropy loss as coarse-graining integrates out internal degrees of freedom, shifting free-energy minima so that the system may exhibit a premature phase transition to a frozen state, even at conditions where the original atomistic system is liquid\cite{sokhan2023,Kidder2021}.  While our \nDPD\ approach elegantly tackles the phase coexistence problem of pure DPD liquids, there is still an inevitable loss of entropy due to isotropic coarse-graining of the internal liquid structure, especially if this is influenced by external factors such as the presence of interfaces and external electromagnetic fields.  By polarising CG liquid models one could thus alleviate the effects of entropy loss, including artificial crystallation or freezing at lower temperature\cite{yss+10}, by representing internal structure re-arrangements as a local dielectric reaction.

This work demonstrates the development of polarisable water models for a comparatively high CG level of 5 molecules per entity, intended for initial use in aqueous electrolytes found in vanadium redox flow batteries\cite{boccardo2024} but also more generally applicable for mesoscopic simulations.  Starting from an existing non-polar \nDPD\ model for water at this CG level\cite{seaton2024}, we attach two charged\cite{wva+13,wv14} sites -- see Figure~\ref{3sm} \emph{left} and \emph{middle} cases -- to the centre of each water bead to form dipole moments from each three-site molecule-like entity, using charge-smearing to avoid the risk of charge collapse.  In each case, each central bead interacts with others via an \nDPD\ potential, while the attached charges only interact electrostatically with the rest of the medium and, when not using rigid bodies, via a harmonic bond and an angle potential with their central beads.  

We initially explored two approaches when selecting charges, both shown in Figure~\ref{3sm}: (1) a two-charge model\cite{vjt18,chiacchiera2024} (2CM) with a neutral central bead and equal magnitude/opposite sign charges on the attached particles ($\pm q$), and (2) a three-charge model\cite{wu2010} (3CM) with equal magnitude/sign charges on the attached particles ($+q$) and a larger opposite sign charge on the central bead ($-2q$) to make the entities electro-neutral. The selected charge magnitude $q$ goes hand-in-hand with the background medium permittivity (selected via Bjerrum length $\lambda_B$ as vacuum) and the bond interactions between the central and outer beads: we have followed a previously set-out approach to determine these parameters\cite{chiacchiera2024}.  We have also additionally explored constraining our model to average bond lengths and angles (via an angle potential) as well as completely rigidifying the entities and replacing flexible bond and angle potentials. %, as well as the use of computationally cheaper methods of obtaining Coulombic interactions between charged beads such as damped shifted and truncated pairwise forces\cite{fennell2006} and the reaction field approach\cite{neumann1986}.  
For the sake of simplicity and computational efficiency in simulations, the third option given in Figure~\ref{3sm} of a three-site model with constant point dipole and quadrupole moments (3DM) was omitted.

In our preliminary studies on the 2CM and 3CM approaches, we started with our non-polar \nDPD\ model and performed a simulation-based systematic search over the parameter space of test charges $q$ and balancing force constants $\kappa$ to evaluate the models' numerical stability and capability to achieve a water-like dielectric response. Results with test charges in the range of $0.75e^{-}~\leq~q~\leq~1.25e^{-}$ were found to be the most promising, leading to two design decisions similar in spirit to those by Wu \emph{et al.}\cite{wu2010}: (i) setting the model test charge to $1e^-$, and (ii) continuing with the 3CM approach. The 3CM approach was ultimately favoured since it generates double the polarisation per linear distance in each entity compared with 2CM, it naturally generates a quadrupole moment while the 2CM does not and, last but not least, simulation-wise it proved to offer better numerical stability than the 2CM, which requires a fine balance between the charge-charge attraction and the rest of the model's force field.

Each model was subjected to two sets of checks in order to arrive at a finely-tuned polarised \nDPD\ water model.  The first set of checks examined thermodynamic and hydrodynamic properties of the polarisable water -- including vapour-liquid coexistence densities, its diffusivity and shear viscosity -- and compared these with those obtained from the original non-polar \nDPD\ water model\cite{seaton2024}, making any required changes to parameters to fit these properties to experimental values.  The second set compared the dipole and quadrupole moments measured using our polarisable water model with those obtained by clustering atomistic water molecules. The clustering of water molecules used both topologically-consistent and purely (volumetrically minimal) stoichiometric approaches to select the atomistic contents of water molecules, respectively with and without regard for molecular bonding connectivity.

\section{Model setup}

\subsection{Design principles}

We build our polarisable water models by furnishing our original solvent `beads' with two Drude sites and preserving the `charge-neutrality' of the new assembly, as seen in Figure~\ref{3sm} \emph{left} and \emph{middle}, with the intent to minimally perturb the \nDPD\ interactions used between the central solvent beads.  The \nDPD\ interactions\cite{sokhan2023} are described by the pairwise interaction potential:
\begin{equation}
\label{eq:nDPDpotential}
U^{n} (r_{ij}) = \frac{A_{ij} b_{ij} r_c}{n+1} \left(1-\frac{r_{ij}}{r_c} \right)^{n+1} - \frac{A_{ij} r_c}{2} \left(1-\frac{r_{ij}}{r_c} \right)^{2}  
\end{equation}
for inter-particle distances $r_{ij}$ less than a cutoff $r_c$, so $U^{n} = 0$ for $r_{ij} \ge r_c$.  In the above, $A_{ij}$ is the repulsive parameter with units of force, $b_{ij}$ is a scaling factor defining the magnitude of repulsion to attraction, and $n$ is a repulsive exponent.  In order for this potential to have both attraction and repulsion, $b_{ij}$ needs to be greater than $\tfrac{1}{2} (n+1)$. The root of this potential $0 < \sigma \le r_c$ defines the length scale for our calculations, given as
\begin{equation}
\label{eq:nDPDsigma}
\frac{\sigma}{r_c} = 1 - \left(\frac{n+1}{2b_{ij}}\right)^{\frac{1}{n-1}} .
\end{equation}
and is therefore determined from the selected values of $n$ and $b_{ij}$.  The resulting conservative force acting between central beads is related to the derivative of Eq. \ref{eq:nDPDpotential} with respect to $r_{ij}$, i.e.
\begin{equation}
\label{eq:nDPDforce}
\vec{F}^{C,n}_{ij} = A_{ij} \left[b_{ij} \left(1 - \frac{r_{ij}}{r_c} \right)^n - \left(1 - \frac{r_{ij}}{r_c} \right) \right] \frac{\vec{r}_{ij}}{r_{ij}} ,
\end{equation}
where $\vec{r}_{ij} = \vec{r}_{j} - \vec{r}_{i}$ is the vector between particles $i$ and $j$. Setting this force to zero gives the length scale for the minimum potential $r_{min}$, equal to
\begin{equation}
\frac{r_{min}}{r_c} = 1 - b_{ij}^{-\frac{1}{n-1}}
\end{equation}
and a minimum potential energy of 
\begin{equation}
U^{n}_{min} = \frac{A_{ij} r_c}{2 b_{ij}^{\frac{2}{n-1}}} \left(\frac{2}{(n+1) b_{ij}^{\frac{n-2}{n-1}}} -1 \right) .
\end{equation}
It should be noted that setting $n = 1$ and $b_{ij}=2$ produces `standard DPD', i.e. purely repulsive Groot-Warren interactions\cite{gw97}, while thermodynamic stability for \nDPD\ interactions is obtained when
\begin{equation}
\label{eq:nDPDstabile}
b_{ij} > \frac{(n+1)(n+2)(n+3)(n+4)}{120}
\end{equation}
and the total configurational energy of the system is bounded from below.  We note that the choices of $n$ and $b_{ij}$ affect the thermodynamics of our solvent beads: the value of $n$ has an effect on vapour-liquid coexistence, particularly the variation of liquid density with temperature, while $b_{ij}$ affects the interfacial tension between these phases and can be selected to give an appropriate trend of this property with temperature for our given solvent (water).  The CG level for our solvent provides the required cutoff distance $r_c$, also affecting our calculation length scale $\sigma$ with $b_{ij}$ and $n$ (see Eq. \ref{eq:nDPDsigma}), while the value of $A_{ij}$ predominately affects the critical temperature for the solvent and can be selected to provide the required reduced temperature (temperature relative to its critical value) and thus the appropriate liquid or vapour densities\cite{seaton2024}.  We will discuss later how this value needed to be adjusted for our CG polarisable water models.

By default, two satellite Drude particles are attached to the central bead of each CG entity, interacting via a harmonic spring potential:
\begin{equation}
U^{bond} = \frac{\kappa}{2} \left(r_{ij} - r_0\right)^2 ,
\label{eq:HarmSpring}
\end{equation}
where $\kappa$ is the harmonic bond force constant and $r_0$ is the chosen equilibrium bond length.  The value of $\kappa$ strongly influences the maximum possible simulation timestep $\Delta t$ that can be used to integrate forces on particles without calculation instabilities.  We note that lower values of $\kappa$ (and thus larger $\Delta t$) can be used when $r_0$ is set to zero, and these naturally lead to broader distributions in bond length and thus dipole moments.  We can further influence the distances between partial charges by applying a harmonic angle potential between the entity's internal pseudo-bonds:
\begin{equation}
U^{angle} = \frac{\kappa_{\theta}}{2} \left(\theta_{ijk} - \theta_0\right)^{2} ,
\label{eq:HarmAngle}
\end{equation}
where $\kappa_{\theta}$ is the harmonic angle potential constant, $\theta_{ijk}$ is the angle formed between the two bonds and $\theta_0$ is the equilibrium angle.

The flexibility of the bonds and angles contributes to the \emph{polarisability} of our water model, enabling the dipole moments (and any quadrupole moments) to vary when subjected to external electric fields or discontinuities in dielectric at interfaces.  We do note, however, that even if the relative positions of the partial charges remain constant as a result of applying rigid body (RB) dynamics instead of harmonic bond and angle potentials, the molecule-like entities still respond to external fields by e.g. rotating and reorientating their charges as well as via inter-bead charge penetration due to the natural softness of the \nDPD\ interactions.  As such, we additionally derive a rigidified model, still providing a \emph{polar} water model but with the benefit of somewhat lower computational cost to solution as the timestep size can be maximised when bonds vibrations are disregarded. 

The interactions between charges are generally defined as a modified Coulombic potential\cite{gmv+06}:
\begin{equation}
\label{smearCoulomb}
U^{e} = \frac{\lambda_B Z_i Z_j}{r_{ij}} \left[1 - f(r_{ij}) \right] ,
\end{equation}
where $\lambda_B = \frac{e^2}{4\pi \epsilon k_B T}$ is the Bjerrum length that inversely scales with the permittivity of the background medium $\epsilon=\epsilon_{r,b}\epsilon_0$, $Z_i=q_i/e$ and $Z_j=q_j/e$ are the charge valencies on particles $i$ and $j$, and $f (r_{ij})$ is a distance-dependent smearing function used to prevent opposite-sign charged particles from permanently collapsing on top of each other.  For our water models, we set the background medium to be vacuum and thus its relative permittivity $\epsilon_{r,b} = 1$ (i.e. $\epsilon = \epsilon_0$ is used to calculate $\lambda_B$). 

The usual approach to solve Eq. \ref{smearCoulomb} is to use Ewald summation, dividing the potential between a short-range pairwise part carried out in real space up to a cutoff distance and a long-range part in reciprocal or Fourier space, often solved using techniques such as Smooth Particle Mesh Ewald (SPME)\cite{essmann1995}.  The real space part of the Ewald sum is modified to include charge smearing, i.e.
\begin{equation}
U^{e,real} = \frac{\lambda_B Z_i Z_j}{r_{ij}} \left[\textrm{erfc} \left(\alpha r_{ij}\right) - f(r_{ij}) \right] ,
\end{equation}
where $\alpha$ is a Gaussian charge-screening parameter that, along with the real space cutoff distance, influences the truncation error in this potential.  The charge smearing function $f (r_{ij})$ is usually chosen to recover the standard Coulombic potential within a small tolerance at or before the real space cutoff, meaning no modification to the reciprocal space part of the calculation is necessary.  We go further here by selecting a Gaussian charge smearing function\cite{wva+13,wv14}:
\begin{equation}
f(r_{ij}) = \textrm{erfc} \left(\frac{r_{ij}}{2 \sigma_G}\right) ,
\end{equation}
where $\sigma_G$ is a smearing length.  Setting $\alpha = \frac{1}{2\sigma_G}$, or \emph{vice versa}, reduces all Ewald real space terms to zero and thus the calculations can be carried out entirely in reciprocal space only, thus saving on the computation load.

Other than electrostatically, the Drude particles and the central beads in the CG water entities do not interact with one another classically, with the exception of the harmonic spring and/or angle potentials when these are within the same CG assembly.
However, when the Galilean-invariant DPD thermostat\cite{gw97} is applied, additional sets of dissipative and random pairwise forces are applied between \emph{all} particles (central beads and Drude particles alike), which are used to control the simulation temperature while conserving momentum:
\begin{equation}
\label{eq:DPDdiss}
\vec{F}^{D} = -\gamma_{ij} \left[w (r_{ij}) \right]^2 \left(\vec{r}_{ij} \cdot \vec{v}_{ij} \right) \frac{\vec{r}_{ij}}{r_{ij}^2} ,
\end{equation}
\begin{equation}
\label{eq:DPDrand}
\vec{F}^{R} = \sqrt{2k_B T \gamma_{ij}} w (r_{ij}) \frac{\xi_{ij}}{\sqrt{\Delta t}} \frac{\vec{r}_{ij}}{r_{ij}} .
\end{equation}
These thermostatting forces depend upon the relative velocity for the particle pair, $\vec{v}_{ij} = \vec{v}_{j} - \vec{v}_{i}$, a switching function $w (r_{ij})$, Gaussian random numbers with zero mean and unity variance for each pair $\xi_{ij}$ and a dissipative force parameter $\gamma_{ij}$.  The transport properties of the entities, including their diffusivities and viscosities, depend on both their interactions and the functional forms of the dissipative and random forces\cite{marsh1997}: since the interactions are fixed to provide correct thermodynamic behaviour and relative permittivity, we can vary $w (r_{ij})$ and values of $\gamma_{ij}$ to match up the diffusivity and/or viscosity to experimental values.  For the sake of simplicity, we set $w (r_{ij}) = 1 - \frac{r_{ij}}{r_c}$ for $r_{ij} < r_c$ and use the same value of $\gamma_{ij}$ for all particle pairs.  We set $r_{c}$ for the dissipative and random forces between pairs of central beads to the same value as for \nDPD\ interactions as in the original non-polar model, but use $r_{c} = \sigma$ for all other particle pairs to limit the range of dynamic thermostatting of the Drude particles to that of their immediate environment.

It is worth noting that the dynamics as well as the transport properties can also be influenced further by the internal mass distribution within the bead.  While several previous studies\cite{wva+13,wv14,wu2010} have chosen $m_{c}/m_{s} = 2$, we decided to increase this ratio to $m_{c}/m_{s} = 4$ to bring the centre-of-mass closer to the central bead, as in Vaiwala \emph{et al.}\cite{vjt18}.

In this study, we were able to apply a modified version of the DPD code in \DLMESO\cite{sam+13} that incorporated the \nDPD\ interaction model and \DLPOLY\cite{dlp5} for simulations of our non-polar and polarisable water models. Both codes are capable of modelling DPD/\nDPD-based polarisable water models with smeared charges and harmonic bond and angle interactions between central and Drude particles, producing statistically identical results for such systems. For rigid body water models, we made exclusive use of \DLPOLY\ with its implementation of the NOSQUISH\cite{miller2002} rigid body dynamics solver.

\subsection{Relative permittivity, dipole and quadrupole moments}

The primary property we aimed to achieve for our polarisable water models was the correct permittivity in water compared with vacuum, $\epsilon_r \approx 78.4$. For a system with periodic boundary conditions and no externally applied electric field, this can be calculated from the global (box) dipole moment\cite{fs02}:
\begin{equation}
\label{dielec}
\epsilon_r = 1 + \frac{\langle P^2\rangle}{3 \epsilon_0 V k_B T} = 1 + \frac{4 \pi \lambda_B \langle P^2 \rangle}{3e^2V} ,
\end{equation}
where $\epsilon_0$ is the permittivity in vacuum, and $\vec{P} = \sum_{i}^{\Nm} \vec{p}_i$, i.e. the sum of all molecular dipole moments $\vec{p}_i$. The dipole moment of molecule $i$ is calculated from the charges and positions for each of its particles (adjusting the latter if the molecule crosses any periodic boundaries):
\begin{equation}
\vec{p}_i = \sum_j^N q_j \vec{r}_j .
\end{equation}
The choice of particle charges $q_j$ and the distances between pairs of charges, either explicitly chosen in rigid bodies or governed by bond and angle interactions, ultimately determines the relative permittivity of polarisable water via Eq. \ref{dielec} \cite{chiacchiera2024}. 

The molecular quadrupole moment tensor $\hat{Q}_i$ can also be determined from its contituent particles' charges and positions relative to the coordinate system origin.  For component $\alpha \beta$, where $\alpha$ and $\beta$ can each be Cartesian coordinates $x$, $y$ or $z$:
\begin{equation}
\label{qab}
Q_{i,\alpha\beta} = \sum_j^N q_j \left(3 r_{j,\alpha} r_{j,\beta}  - r_j^2 ,\delta_{\alpha\beta}\right)
\end{equation}
where $\delta_{\alpha\beta}$ is the Kronecker delta equal to 1 when $\alpha = \beta$ and 0 when $\alpha \neq \beta$.  The quadrupole moment tensor is traceless by construction, $Q_{xx} + Q_{yy} + Q_{zz} = 0$, although the separate components $Q_{\alpha\alpha}$ are likely to be non-zero as are the amplitudes $|Q_{\alpha}|=\sqrt{\left(Q_{\alpha x} + Q_{\alpha y} + Q_{\alpha z} \right)}$ for charged entities as three-charge models (3CMs).  It is worth noting that calculation of $\hat{Q}$ can be carried out in any frame of reference: in the internal natural frame of reference of an atomistic water molecule $\hat{Q}$ is a diagonal matrix\cite{niu2011}.

In the analyses that follow, we depict ensemble- and time-averaged properties such as dipole and quadrupole moments with angular brackets, e.g. $\langle \vec{p} \rangle$, while spatial averages over Cartesian dimensions are shown with lines above the property in question, e.g. $\overline{Q_{\alpha}} = \tfrac{1}{3} \left(Q_{x} + Q_{y} + Q_{z}\right)$.  It is worth noting that the reported values of these properties are also averaged per CG entity: multiplying these quantities by the number density of the CG entities gives the system-wide (macroscopic) values of the properties per unit volume.

\begin{figure}
    \centering
    \includegraphics[width=\columnwidth]{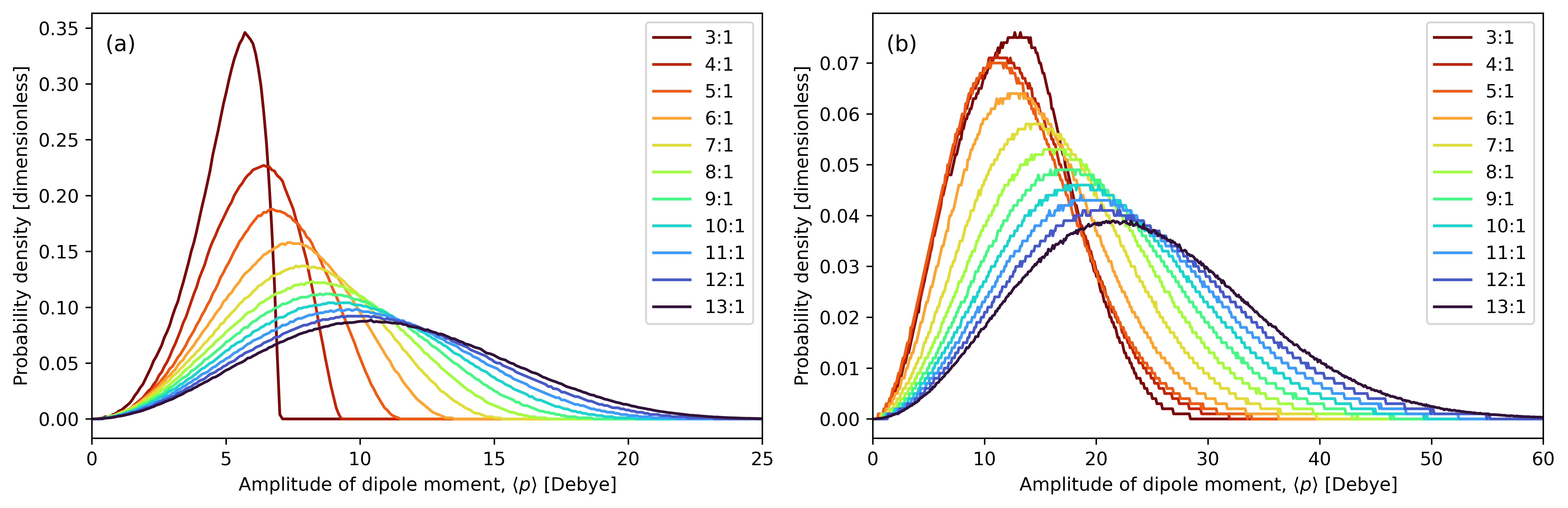}
    \caption{Averaged per-cluster dipole moment amplitude probability distributions derived from TIP3P water simulations under normal conditions at different levels of coarse-graining -- clusters of 3-13 water molecules per bead, using two different methods for coarse-graining sampling -- (a) topological and (b) stoichiometric as described in the text.}
    \label{add}
\end{figure}

\begin{figure}
    \centering
    \includegraphics[width=\columnwidth]{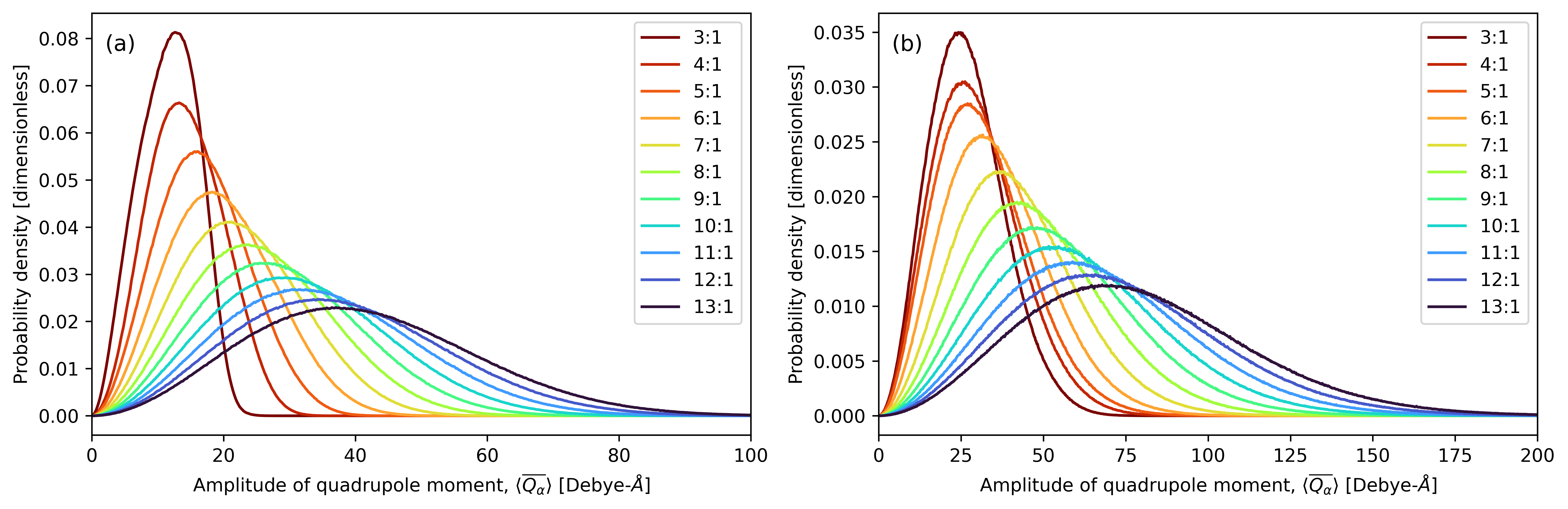}
    \caption{Averaged quadrupole moment amplitude (calculated in the global coordinate system) probability distributions derived from TIP3P water simulations under normal conditions at different levels of coarse-graining -- clusters of 3-13 water molecules per bead, using two different methods for coarse-graining sampling -- (a) topological and (b) stoichiometric as described in the text.}
    \label{aqd}
\end{figure}

\begin{figure}
    \centering
    \includegraphics[width=\columnwidth]{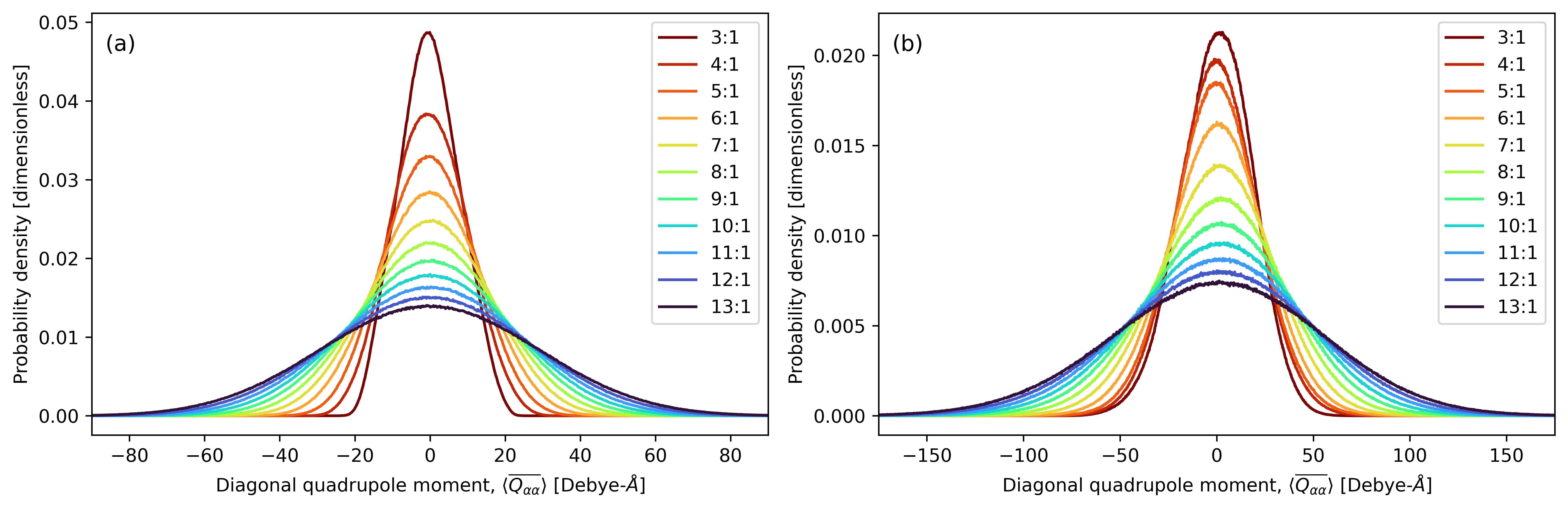}
    \caption{Averaged diagonal component of quadrupole moment (calculated in the global coordinate system) probability distributions derived from TIP3P water simulations under normal conditions at different levels of coarse-graining -- clusters of 3-13 water molecules per bead, using two different methods for coarse-graining sampling -- (a) topological and (b) stoichiometric as described in the text.}
    \label{aqdi}
\end{figure}

\section{Connection to Molecular Dynamics water}

To validate the goodness of our polarised CG water models and connect them to `first principles' atomistic water models, we chose a system of 7,263 water molecules based on the default rigid-body TIP3P model \cite{klein_83} and carried out simulations using \DLPOLY \cite{dlp5}.  Following best practices in modelling and simulation, a $300$~ps equilibration using the NPT (Martyna-Tuckerman-Klein, $\tau_t = 1.0$~ps, $\tau_p = 5.0$~ps) ensemble at $T = 295$~ K and $p = 1$~atm with velocity rescaling was carried out before collecting 1,000 trajectory frames of a $100$~ps simulation in the NVT (Nos\'{e}-Hoover, $\tau_t = 1.0$~ps) ensemble.

We processed the trajectory to calculate the dipole and quadrupole moments distributions at different levels of coarse-graining, $l_{\textbf{CG}}$, the value of which equals the equivalent number of water molecules contained in a cluster.  We built these distributions using two distinctively different methods for CG sampling of the MD cell bulk, both of which are valid from a mesoscopic point of view.  The first method, as outlined in Wu \emph{et al.}\cite{wu2010} and referred to by us as `topologically consistent', traversed all distinctively different and minimal clusters of atomistic water molecules for the level of CG while passing; in this context, a CG level of $l_{\textbf{CG}}$ refers to clusters of the $l_{\textbf{CG}}-1$ closest and topologically distinctive water molecules around each molecule in the simulated system.  The second method, which we refer to as 'purely stoichiometric', disregarded  molecular bonding connectivity and traversed the closest in distance set of atoms to a central molecule to conserve the molecular substance stoichiometry for the CG level.  For example, at a CG level of $l_{\textbf{CG}}$ we sit on the oxygen of a central molecule and consider only the closest set of oxygen and hydrogen atoms that conserves the stoichiometry: this set therefore includes the closest $2l_{\textbf{CG}}$ hydrogen and $l_{\textbf{CG}}-1$ oxygen atoms.

Figure~\ref{add} shows how the dipole moment magnitude distribution for each cluster, $\langle p \rangle$, varies with the CG level for our chosen atomistic water model.  The left and right panels of the figure exhibit similar trends for the distributions, including how they spread to higher values and become somewhat more symmetric for higher CG levels, although they also show how markedly they differ for the same CG levels when different coarse-graining methodologies are applied.  This supports the \emph{dualistic nature} of mesoscopic modelling: merging together atomistic and engineering modelling scopes while also implying both greater flexibility in designing a model with respect to a desired property and a larger reproducibility space for any specific property. e.g. dipole moments distribution, that a mesoscopic model should also aim to include. Since both representations in Figure~\ref{add} are valid, our models should therefore attempt to represent features from both.

It is worth noting that our results for the 3:1, 4:1 and 5:1 topological dipole moment distributions agree closely with those shown in Wu \emph{et al.}\cite{wu2010} (4:1) and Riniker \emph{et al.}\cite{riniker2011} (3:1, 4:1 and 5:1) respectively.  The generated distributions also varied very little when we substituted our chosen atomistic water model with other popular ones, e.g. TIP3P-F (flexible TIP3P), TIP4P and SPC/E.  This gave us the confidence that all of these models, while subtly different, contain the essential core information about the medium dielectric and polarisability that we could pass up the length scales to build into our mesoscopic models.

Figure~\ref{aqd} shows the averaged quadrupole moment amplitude distributions calculated in the global coordinate system, $\langle\overline{Q_{\alpha}}\rangle$, and their dependence on the CG level using the same water core model (TIP3P) and set of data.  Again, our results for the 4:1 topological quadrupole moment amplitude distribution agree well with those shown in Wu \emph{et al.}\cite{wu2010}. Figure~\ref{aqdi} shows the distributions of the average diagonal component of the quadrupole moment calculated in the global coordinate system, $\langle\overline{Q_{\alpha\alpha}}\rangle$, and their dependence on the CG level using the same water core model (TIP3P) and set of data.  Our results for the 5:1 topological quadrupole moment diagonal distribution exhibits the correct behaviour as shown in Li \emph{et al.}\cite{li2020}, although it is broader by a factor of approximately 3.

\begin{figure}
    \centering
    \includegraphics[width=\columnwidth]{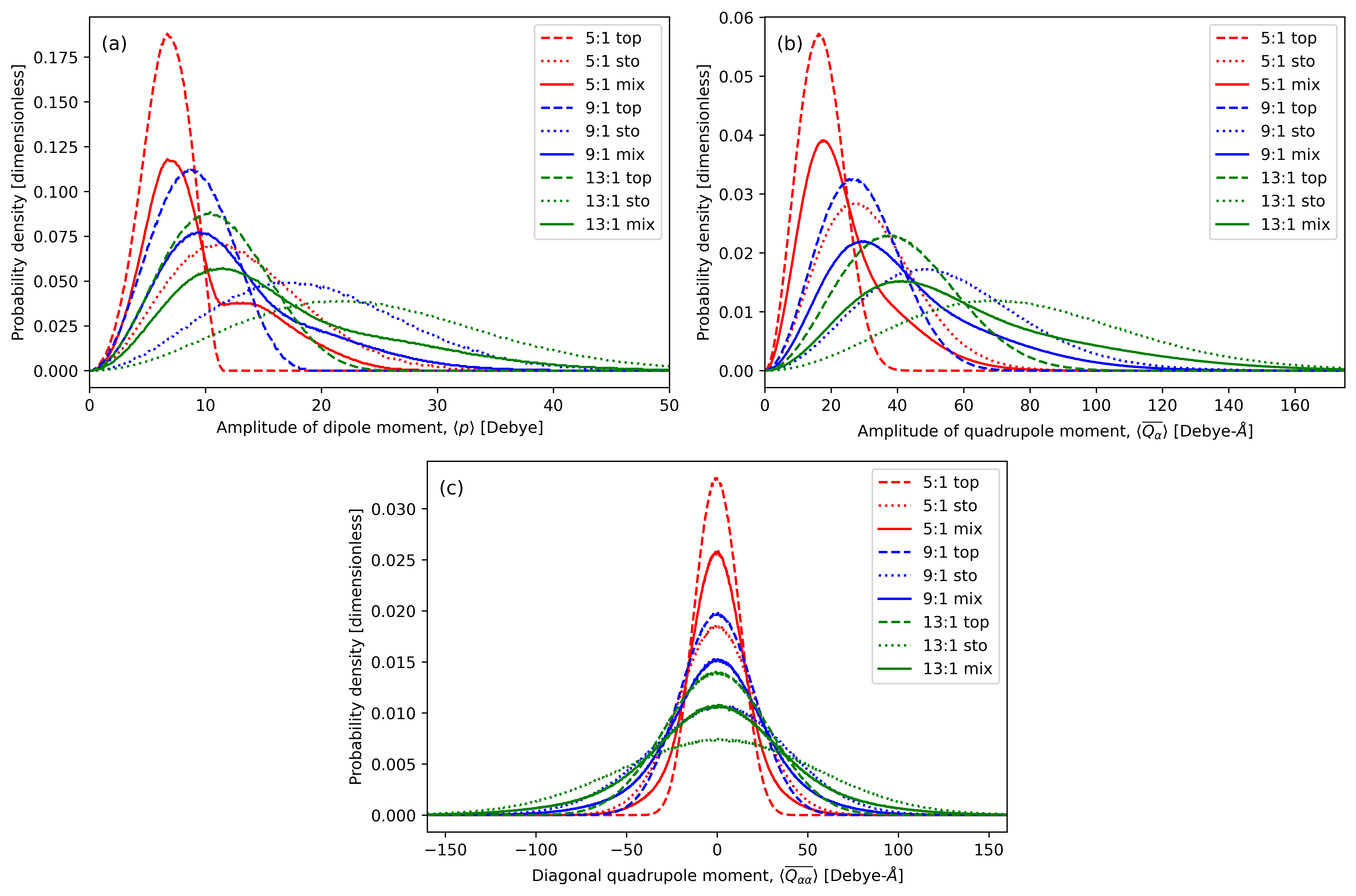}
    \caption{Probability distributions of (a) dipole moment magnitude, (b) average quadrupole moment amplitude and (c) average diagonal component of quadrupole moment (calculated in the global coordinate system), obtained by \textbf{mix}ing those derived by coarse-graining from TIP3P water model simulations for clusters of 5, 9 and 13 water molecules per bead, using \textbf{top}ological and \textbf{sto}ichiometric approaches described in the text.}
    \label{amd}
\end{figure}

Figure~\ref{amd} is a composite of the previous figures that compares and contrasts the changes in probability distributions of $\langle p \rangle$, $\langle\overline{Q_{\alpha}}\rangle$ and $\langle\overline{Q_{\alpha\alpha}}\rangle$ for key CG levels of 5:1, 9:1 and 13:1 with three different types of coarse-graining, based on the same water core model (TIP3P) and set of atomistic trajectory data.  Here, we have introduced an additional \textbf{mix}ed type of coarse-graining that is a 50/50 mix of the topological and stoichiometric approaches.  The 50/50 ratio was picked in the spirit of the mesoscopic scale to represent both approaches equally.  The new CG distributions inherit the broadness of the stoichiometric ones and, for $\langle p \rangle$ and $\langle\overline{Q_{\alpha}}\rangle$, shift their maxima $Max(\langle p \rangle)$, $Max(\langle\overline{Q_{\alpha}}\rangle)$ and expected (average) values $E(\langle p \rangle)$, $E(\langle\overline{Q_{\alpha}}\rangle)$ to somewhat higher values than those obtained using the topological CG approach.

Increasing the CG level, $l_{\textbf{CG}}$, for a system as interesting as liquid water provides an opportunity to better average out the atomistic representations of the liquid, and thus improve how the dielectric nature of the polar liquid is embedded; this can be represented by microscopically-driven internal properties such as $\langle p \rangle$ and $\langle\hat{Q}\rangle$ within the beads that increase along with the CG level.  Hydrogen bonding is a crucial molecular feature of water, one over which we should take care when averaging out and is important to represent efficiently across CG levels.  Given its electrostatic nature, its effective inclusion will affect the average CG entity's internal dipole and quadrupole moments, thus directly influencing the macroscopic dielectric response of the new CG medium, including its ability to polarise.  It is worth noting that as far as hydrogen bonding is concerned, the difference between the topological and stoichiometric CG averaging methods starts to matter, especially at lower levels (excluding 2:1) where the former method is somewhat less inclusive of hydrogen bonding, whereas the latter favours it by more efficient inclusion arising from stoichiometric volume averaging.  The relative difference between the two CG methods could therefore provide an answer to the question of identifying a suitable coarse-graining level for our liquid of choice.  

We define the relative difference of a polarisable property $b$ of a liquid calculated between the two different CG schemes as $\{b\}~=~E\left(\langle b_{\bf sto}~-~b_{\bf top}\rangle\right)/E\left(\langle b_{\bf mix}\rangle\right)$, where $\langle b_{\bf mix}\rangle~=~\langle b_{\bf sto}~+~b_{\bf top} \rangle/2$.  Equipped with this notation, we can examine the behaviour of $\{p\}$, $\{\overline{Q_\alpha}\}$, $\{p.p\}$, $\{Q:Q\}$ and $\{Q:Q/p.p\}$ across different CG levels, $l_{\textbf{CG}}$, as given in Table~\ref{tab:cgl}.  Excluding the CG level of 2:1 ($l_{{\textbf{CG}}}=2$) for which $\{p\}$ and $\{\overline{Q_\alpha}\}$ are very small, we can readily conclude that a CG level of 5:1 (five molecules of water per CG bead) is very much justifiable, since it offers the smallest deviations (relative differences) between the two CG representations for $\{p\}$, $\{p.p\}$ and $\{Q:Q/p.p\}$.  While a CG level of 6:1 offers more of an advantage in the cases of $\{\overline{Q_\alpha}\}$ and $\{Q:Q\}$, we note that our chosen CG level of 5:1 still offers an excellent improvement over the next level down (4:1).  The additional properties in Table~\ref{tab:cgl} beyond the bead's internal dipole moment $p$ and quadrupole moment $\hat{Q}$ are important\cite{Onsager1936,Fernandez1997,Dimitrova2020} since the microscopic (dipole-dipole) polarisability (polar strength) relates to $p$ via $\alpha_p\propto~p.p$, while the microscopic quadrupolarisability (dipole-quadrupole polarisability or quadrupole strength) $\alpha_Q$ is related to $\hat{Q}$ as $\alpha_Q\propto~Q:Q$.  The quadrupole strength $\alpha_Q$ is also related to another characteristic of the continuous medium, $L_Q$ -- the associated quadrupole length of the dielectric liquid -- as $\alpha_Q=3\epsilon_0\epsilon_r~L_Q^2$, making $L_Q^2\propto~Q:Q/p.p$.

\begin{table}
  \centering
%  \begin{ruledtabular}
    \begin{tabular}{lllllll}
      $l_{\textbf{CG}}$ & $\{p\}$ & $\{\overline{Q_{\alpha}}\}$ & $\{p.p\}$ & $\{Q:Q\}$ & $\left\{\frac{Q:Q}{p.p}\right\}$ & $\delta L_Q (\%)$ \\
      \hline\\[-9pt]
      2	  & 0.956 & 0.203 & 1.566 & 1.404 & -0.418 & 8.46 \\
      3	  & \textit{0.878} & \textit{0.789} & \textit{1.508} & \textit{1.383} & -0.291 & \textit{8.23} \\
      4	  & 0.758 & 0.696 & 1.373 & 1.267 & -0.201 & 7.18 \\
      5	  & \textbf{0.657} & 0.606 & \textbf{1.242} & 1.138 & \textbf{-0.170} & 6.42 \\
      6	  & 0.669 & \textbf{0.579} & 1.244 & \textbf{1.090} & -0.251 & 4.89 \\
      7	  & 0.701 & 0.586 & 1.277 & 1.098 & -0.302 & \textbf{4.25} \\
      8	  & 0.722 & 0.600 & 1.300 & 1.116 & -0.317 & 4.34 \\
      9	  & 0.734 & 0.611 & 1.312 & 1.132 & -0.317 & 4.64 \\
      10  & 0.741 & 0.618 & 1.319 & 1.141 & -0.314 & 4.95 \\
      11  & 0.746 & 0.622 & 1.324 & 1.147 & -0.315 & 5.14 \\
      12  & 0.750 & 0.6237 & 1.327 & 1.1483 & -0.320 & 5.21 \\
      13  & 0.754 & 0.6244 & 1.331 & 1.1482 & \textit{-0.328} & 5.16 \\ [-1pt]
    \end{tabular}
%  \end{ruledtabular}
  \caption{Dimensionless relative differences of core polarisable properties, $\{b\}~=~E\left(\langle b_{\bf sto}~-~b_{\bf top}\rangle\right)/E\left(\langle b_{\bf mix}\rangle\right)$, for water between the two different CG schemes, {\bf sto}ichiometric and {\bf top}ological, at different CG levels, $l_\textbf{CG}$, derived from the same molecular dynamics trajectory based on the TIP3P water model.  The smallest value in magnitude of each evaluated property across the $l_{\textbf{CG}}$ series, excluding those for a water dimer ($l_{{\textbf{CG}}}=2$), is indicated in \textbf{bold}, whilst the largest is \textit{italicised}. \label{tab:cgl}}
\end{table}

All levels of coarse-graining, $l_{\textbf{CG}}$, should ideally reproduce the same quadrupole strength per molecule of water, i.e. $\alpha_{Q,{l_\textbf{CG}}} = \alpha_{Q,l_{\textbf{CG}}=1} \cdot l_\textbf{CG}^{1/3}$, where $\alpha_{Q,l_{\textbf{CG}}=1}$ is the quadrupole strength for a single water molecule and the molecular or bead density for the liquid is assumed to be constant. We can therefore relate the bead's quadrupole length averaged per molecule, $L_{Q,{l_{\bf{CG}}}}^\textbf{mol}$, to that of the atomistic model, $L_{Q,{l_{\textbf{CG}}=1}}$ via:
\begin{equation}
L_{Q,{l_{\bf{CG}}}}^{\bf{mol}}~=~\left( \frac{\langle Q:Q/p.p \rangle_{l_{\textbf{CG}}}~/~{{l_{\textbf{CG}}}^{1/3}}}{\langle Q:Q/p.p \rangle_{l_{\textbf{CG}}=1}} \right)^{1/2}~L_{Q,l_{\textbf{CG}}=1}
\end{equation}
in which $L_{Q,{l_{\textbf{CG}}=1}}$ can be equated to $L_Q$ for water, typically between $1$ and $2$ \AA.  What is of greater interest here, however, is evaluating the effect the coarse-graining level has on the relative quadrupole length per molecule and how much this deviates from the expected value. We therefore define the relative deviation (or compromise) as $\delta L_Q(l_{\textbf{CG}})=L_{Q,{l_{\bf{CG}}}}^{\bf{mol}}/L_Q - 1$ and observe how this changes as $l_{\textbf{CG}}$ increases and more and more degrees of freedom are integrated out dynamically by clumping together and averaging out the water molecules electrostatically into CG beads.  As seen in Table~\ref{tab:cgl}, the deviation is immediate and varies across the series of CG levels: a minimum value for this deviation is attained for 7:1, although our final choice of $l_{\textbf{CG}} = 5$ is still tolerable.

\section{Selected models}

Based on the design principles described above, we have arrived at three polarisable water models that produce the correct relative permittivity for water.  The starting point for these water models is a non-polar model with \nDPD\ interactions between beads\cite{seaton2024}, each representing five molecules of water at room temperature (298.15 $K$).  The interaction parameters in this case were set to give the correct liquid density for water at the required reduced temperature (i.e. the ratio of temperature to the critical temperature of water, 647.096 $K$) and an expected temperature dependence on the interfacial tension between co-existing liquid and vapour phases\cite{sokhan2023}.

The parameters for the polarisable water models can be expressed in terms of three selected fundamental quantities: the mass of a water bead ($m$) based on the number of molecules per bead (equal to the coarse-graining level $l_{\textbf{CG}}$), an energy unit based upon temperature ($\epsilon = \kT$), and its radius ($\sigma$) based on $l_{\textbf{CG}}$, the values of $b_{ij}$ and $n$ for the \nDPD\ interactions, Eq. \ref{eq:nDPDsigma}, and matching up the expected bead density with the density of liquid water at the given temperature.  These three quantities were subsequently used as units for DPD-based calculations and can be used to define the units for other quantities such as time, pressure and electric dipoles.  All of the units in use for the water models, including their equivalent `real-life' physical values, are given in Table~\ref{tab:units}.

\begin{table}
  \centering
%  \begin{ruledtabular}
    \begin{tabular}{llll}
      \textrm{Physical quantity}  & DPD unit &
      Example value & unit  \\
      \hline\\[-9pt]
      mass   & $m$    & $1.49575\times10^{-22}$ & kg \\
             &        & $90.0764$ & u (Da) \\
      length & $\sigma$  & $0.623956$ & $\nm$  \\
      energy & $\epsilon = \kT$  & $4.116405\times10^{-21}$ & J \\ 
             &        & $2.47896$ & $\mathrm{kJ}\,\mathrm{mol}^{-1}$ \\
      time   & $\tau = \sigma \sqrt{m/\epsilon}$ & $3.76119$ & ps \\
      electric dipole & $e \sigma$ & $9.99688\times10^{-29}$ & $\mathrm{C}\,\mathrm{m}$ \\
                      &            & $29.9699$ & Debye \\
      electric field  & $\epsilon/(e \sigma)$
      & $4.11769 \times 10^7$ & $\mathrm{V}\,\mathrm{m}^{-1}$\\
      %electric dipole & $e \rc$ & $1.03\times10^{-28}$
      %& $\mathrm{C}\,\mathrm{m}$ \\
      pressure & $\epsilon/\sigma^3$ & 16.9456 & MPa \\
      surface tension & $\epsilon/\sigma^2$ & $10.5733$
      & $\mathrm{mN}\,\mathrm{m}^{-1}$\\[-1pt]
    \end{tabular}
%  \end{ruledtabular}
  \caption{DPD units and physical units for relevant physical quantities.  The values in the third column correspond to the choice $l_{\textbf{CG}}=5$, room temperature ($T=298.15~K$) and matching the bead density to give the equivalent value for liquid water at this temperature.\label{tab:units}}
\end{table}

The three polarisable water models all involve a charged central bead that also interacts with other central beads via the \nDPD\ potential (Eq. \ref{eq:nDPDpotential}) and two attached beads with equal masses and charges (sign and magnitude): the three charges sum up to zero to make the water `super-molecule' charge neutral.  We found these 3CM models to be more numerically stable and easier to parameterise for the correct relative permittivity than 2CM models with two beads of equal mass and charge magnitude (but opposite sign) attached to a neutral central bead.

The distinctions between the models are based upon how the two additional beads are attached to the central bead in each polarisable water entity.  The first model, denoted as \emph{Polar-I}, makes use of a harmonic bond potential (Eq. \ref{eq:HarmSpring}) with the equilibrium bond length $r_0$ set to zero: the strength of the bond $\kappa$ therefore sets its average length and determines the distributions of bond lengths and resulting dipole moments.  The second model \emph{Polar-II} also applies a harmonic bond potential but now sets a non-zero equilibrium bond length $r_0$ to control the (mean) lengths of the two Drude bonds: a higher value of $\kappa$ is now used to control the distribution of bond lengths along with dipole moments\cite{vjt18}, as well as influence the maximum usable value for the simulation timestep $\Delta t$.  This model also adds a harmonic angle potential (Eq. \ref{eq:HarmAngle}) to control the angle between the Drude bonds, also influencing the dipole moment distribution, chosen as the mean angle from \emph{Polar-I}.  The third model \emph{Polar-III} fixes both the bond lengths and the angle between them, treating the entire CG water entity as a rigid body with both translational and rotational dynamics.  The parameters used for all three polarisable models, along with those originally used for the original non-polar \nDPD\ model, are provided in Table~\ref{tab:params}.

\begin{table*}
%  \begin{ruledtabular}
    \begin{tabular}{llrrrr}
      \textrm{Parameter} & Unit & Non-polar & \emph{Polar-I} & \emph{Polar-II} & \emph{Polar-III} \\
      \hline\\[-9pt]
      Mass of central bead, $m_{c}$ & $m$ & 1.0 & 0.666 & 0.666 & 0.666 \\ 
      Charge on central bead, $q_{c}$ & $e$ & 0.0 & -2.0 & -2.0 & -2.0 \\
      Mass of attached bead, $m_{s}$ & $m$ &  & 0.167 & 0.167 & 0.167 \\
      Charge on attached bead, $q_{s}$ & $e$ &  & 1.0 & 1.0 & 1.0 \\
      \nDPD\ central bead repulsion parameter, $A_{ij}$ & $\epsilon \sigma^{-1}$ & 7.56566 & 4.99585 & 4.99585 & 4.99585 \\
      \nDPD\ repulsion/attraction scaling factor, $b_{ij}$ & & 15.0 & 15.0 & 15.0 & 15.0 \\
      \nDPD\ repulsion index, $n$ & & 4 & 4 & 4 & 4 \\
      \nDPD\ interaction cutoff distance, $r_c$ & $\sigma$ & 2.22381 & 2.22381 & 2.22381 & 2.22381 \\
      Harmonic bond strength between beads, $\kappa$ & $\epsilon \sigma^{-2}$ &  & 47.0 & 512.0 &  \\
      Harmonic bond equilibrium distance, $r_0$ & $\sigma$ &  & 0.0 & 0.228 &  \\
      Rigid-body fixed bond length between beads, $l_0$ & $\sigma$ &  &  &  & 0.239 \\
      Harmonic angle strength between bonds, $\kappa_{\theta}$ & $\epsilon$ &  & 0.0 & 10.0 &  \\
      Equilibrium or fixed angle between bonds, $\theta_0$ & $^{\circ} $ &  &  & 96.6 & 96.6 \\
      Bjerrum length of vacuum, $\lambda_B$ & $\sigma$ & 0.0 & 89.8235 & 89.8235 & 89.8235 \\
      Gaussian smearing length for charges, $\sigma_G$ & $\sigma$ &  & 0.517929 & 0.517929 & 0.517929 \\
      Dissipative force parameter, $\gamma_{ij}$ & $\epsilon^{1/2} m^{1/2} \sigma^{-1}$ & $15.5$ & $15.5$ & $15.5$ & $15.5$ \\
    \end{tabular}
%  \end{ruledtabular}
  \caption{Parameters for non-polar and polarisable water models.\label{tab:params}}
\end{table*}

The presence of the attached beads in all three polarisable models changes the thermodynamic behaviour compared with the original non-polar \nDPD\ model, particularly the saturated liquid density. Without changing the \nDPD\ repulsion parameter $A_{ij}$, we observed a 13\% increase in liquid density when transforming the original non-polar \nDPD\ water model into a polar one, most likely due to the addition of charges and the resulting electrostatics-driven clustering of oppositely charged particles. To recover the correct liquid density for water at $298.15~K$, we adjusted the value of $A_{ij}$ downwards for all three polar water models as shown in Table~\ref{tab:params}. Both $A_{ij}$ and the mass of the central bead $m_{c}$ happen to be reduced by approximately one-third from the non-polar model to the polar ones, but since the mass has no influence on thermodynamic behaviour, this is entirely coincidental!

We carried out simulations in NVT ensembles using \DLMESO\ and \DLPOLY\ with the DPD thermostat to confirm the polar water models could obtain correct liquid densities and to observe their other thermodynamic and hydrodynamic properties. Elongated boxes with periodic boundary conditions were used to encourage the CG water entities to separate out into vapour and liquid phases\cite{sokhan2023}; the vapour $\rho_v$ and liquid $\rho_l$ phases could subsequently be measured from the resulting density profiles in the direction of elongation. These vapour-liquid equilibria (VLE) simulations also enabled calculations of interfacial tensions $\gamma$ between the two phases from measured stress tensors.  Equilibrated boxes of water beads or polar entities at the required liquid density were used to calculate self-diffusivities $D$ by measuring time gradients of mean-squared displacements (MSDs). 

We made use of two different methods to obtain the dynamic viscosity $\mu$: (i) sampling off-diagonal compoments of pressure tensors over time from the same equilibrated boxes used for diffusivity calculations and using Einstein-Helfand relations to determine viscosities at zero shear rate\cite{malaspina2023}, and (ii) subjecting boxes to linear shear using Lees-Edwards boundary conditions and calculating ratios of off-diagonal components of the pressure tensor (shear stresses) to velocity gradients (shear rates). While both methods were available in \DLMESO, only the Einstein-Helfand method was available in \DLPOLY\ and, as such, only the zero shear rate viscosity could be determined for \emph{Polar-III}, while both types were available for the non-polar, \emph{Polar-I} and \emph{Polar-II} water models. Finally, Eq.~\ref{dielec} was applied to determine the relative permittivities $\epsilon_r$ from ensemble-averaged squared system dipole moments $\langle P^2 \rangle$, determined using trajectory data from either code.  

\begin{table*}
  \centering
%  \begin{ruledtabular}
    \begin{tabular}{l|r|r|r|r|r|r}
      \textrm{Model} & $\rho_l$/kg~m$^{-3}$ & $\rho_v$/kg~m$^{-3}$ & $\gamma$/$10^{-3}$~N~m$^{-1}$ & $D$/$10^{-9}$~m$^2$~s$^{-1}$ & $\mu$/$10^{-4}$~Pa~s & $\epsilon_r$ \\
      \hline\\[-9pt]
      Non-polar & 1007.0 & 2.59 & 73.0 & $2.50 \pm 0.48$ & $10.33 \pm 0.21$ & 0.0 \\
       & & & & & ($10.63 \pm 1.33$) & \\
      \hline\\[-9pt]
      \emph{Polar-I} & 999.3 & 6.56 & 48.5 & $2.47 \pm 0.48$ & $10.71 \pm 0.27$ & $78.1 \pm 1.4$ \\    
       & & & & & ($10.62 \pm 1.85$) & \\
      \hline\\[-9pt]
      \emph{Polar-II} & 1002.0 & 3.11 & 51.2 & $2.45 \pm 0.71$ & $10.67 \pm 0.18$ & $78.8 \pm 1.4$ \\  
       & & & & & ($10.29 \pm 1.38$) & \\
      \hline\\[-9pt]
      \emph{Polar-III} & 991.3 & 2.01 & 51.9 & $2.46 \pm 0.72$ & - & $77.5 \pm 1.5$ \\
       & & & & & ($10.22 \pm 1.50$) & \\
      \end{tabular}
%  \end{ruledtabular}
  \caption{Properties obtained from non-polar and polarisable water models: liquid density $\rho_l$, vapour density $\rho_v$, interfacial tension $\gamma$, self-diffusivity $D$, dynamic viscosity $\mu$ (zero shear-rate value in brackets) and relative permittivity $\epsilon_r$. These can be compared with experimentally determined values for water at $298.15~K$: $\rho_l = 997.1~\textrm{kg}~\textrm{m}^{-3}$, $\rho_v = 0.023 ~\textrm{kg}~\textrm{m}^{-3}$, $\gamma = 72.0\times10^{-3}~\textrm{N}~\textrm{m}^{-1}$, $D = 2.43\times 10^{-9}~\textrm{m}^2~\textrm{s}^{-1}$, $\mu = 8.891\times 10^{-4}~\textrm{Pa s}$, $\epsilon_r = 78.4$.\label{tab:prop}}
\end{table*}

Table~\ref{tab:prop} confirms our choices of $A_{ij}$ for the original non-polar model and all three polar models, which consistently give a liquid density for water at $298.15~K$ within 1\% of the expected experimentally-determined value. All four models additionally overestimate the saturated vapour density by around two orders of magnitude. Notably, \emph{Polar-I} produces the largest liquid and vapour densities, the latter nearly twice the value predicted by the non-polar and other two models, which suggests that additional clustering of opposite-sign charges is significantly easier to achieve when more flexible bonds are used to attach satellite (Drude) charges to central beads, suggesting that restricting the harmonic bonding via an additional quartic term could be beneficial. Adjusting down the value of $n$ for the \nDPD\ interactions between central beads would modify vapour-liquid equilibrium behaviour, particularly the density ratio $\rho_l/\rho_v$, and it may be possible to do so to obtain improved vapour densities for both non-polar and polar water.  However, doing so will require further adjustment of $A_{ij}$ to accommodate a change in critical temperature resulting from even a small change in $n$ \cite{sokhan2023}.

The interfacial tensions obtained by the polar models all underestimate the expected value by around 30\%, in a similar fashion to a 2CM water model using Lennard-Jones interactions between neutral central beads\cite{riniker2011}. Since the value of $A_{ij}$ was reduced by around a third for all three polar models compared to the original non-polar water model, this indicates that the additional charges do not seem to affect interfacial tension in any significant way. The parameterisation of $b_{ij}$ for the non-polar \nDPD\ water model\cite{seaton2024} made use of this property, so a reduction in this value should increase $\gamma$ towards the experimental value. We note the following caveats in doing so: (i) some care \emph{may} be needed to ensure the condition for thermodynamic stability (Eq.~\ref{eq:nDPDstabile}) is still met, i.e. $b_{ij}$ remains above 14 for our current value of $n=4$, although the existence of bonds to satellite charges might enable further thermodynamic stability; (ii) adjusting $b_{ij}$ may necessitate further changes to $A_{ij}$ to additionally obtain correct liquid densities; and (iii) the value of $b_{ij}$ dictates our length scale $\sigma$, which may require additional modifications to other simulation inputs (e.g. bond strengths and/or equilibrium lengths) if the selected unit system for the simulation code is based upon this quantity. (If simulation inputs are based on `real-life' units such as those in the final column of Table~\ref{tab:units}, this issue can be somewhat circumvented.)

Our choice of dissipative force parameters $\gamma_{ij}$ for all of the water models are justified by good matches of self-diffusivity $D$ to the expected experimentally-determined value. We noted that while the value of $A_{ij}$ can affect self-diffusivity, the required reduction of $A_{ij}$ for the polar models alongside the addition of charge interactions had little to no effect on this property and hence we were able to use the same dissipative force parameter $\gamma_{ij}$ across the non-polar and polar water models. The diffusivities between the three polar models were all broadly similar, certainly within error of measurement via simulations, which suggests that the manner in which the charges are attached have no significant effect on this property. 

The same can be said of the dynamic viscosity $\mu$: the values obtained from the polar water models all match up with each other and are broadly similar to the values obtained for the original non-polar model, all of which use the same value of $\gamma_{ij}$. From our calculations based on applying linear shear, we observe no significant changes in viscosity values as functions of shear rate for the non-polar, \emph{Polar-I} and \emph{Polar-II} models, nor any significant differences with the values obtained at zero shear via equilibrium simulations and pressure tensor autocorrelation functions for all four models: this implies that the non-polar and polar waters all behave as Newtonian fluids. For the given value of $\gamma_{ij}$, all four models overestimate the viscosity by around 15--20\% compared to the experimental value. Since $D$ decreases but $\mu$ increases with increasing $\gamma_{ij}$, any attempt to better match both properties will require modification of the switching function used in the dissipative and random thermostatting forces (Eqs.~\ref{eq:DPDdiss} and \ref{eq:DPDrand}). A viable candidate \cite{fan2006} would be $w (r_{ij}) = \left(1 - \frac{r_{ij}}{r_{c}}\right)^s$ that happens to reduce to the current function when $s=1$: a small modification to this additional parameter would adjust the Schmidt number ($Sc = \frac{\mu}{\rho D}$) to the appropriate value for liquid water ($Sc \approx 367$), thus enabling matches to both diffusivity and viscosity.

We note that the above suggested improvements to the polar models -- lowering $n$ while increasing $A_{ij}$, adjusting down $b_{ij}$ and modifying $w (r_{ij})$ -- are for future studies and not the immediate purpose of this work, i.e. to polarise our extant non-polar \nDPD\ water model and make use of both polar and non-polar models for liquid electrolyte systems, specifically those found in vanadium-based redox flow batteries.  To that end, obtaining the correct liquid density, diffusivity and relative permittivity of water should suffice and thus the current \nDPD-based polar models are adequate for this purpose.

\begin{figure}
    \centering
    \includegraphics[width=\columnwidth]{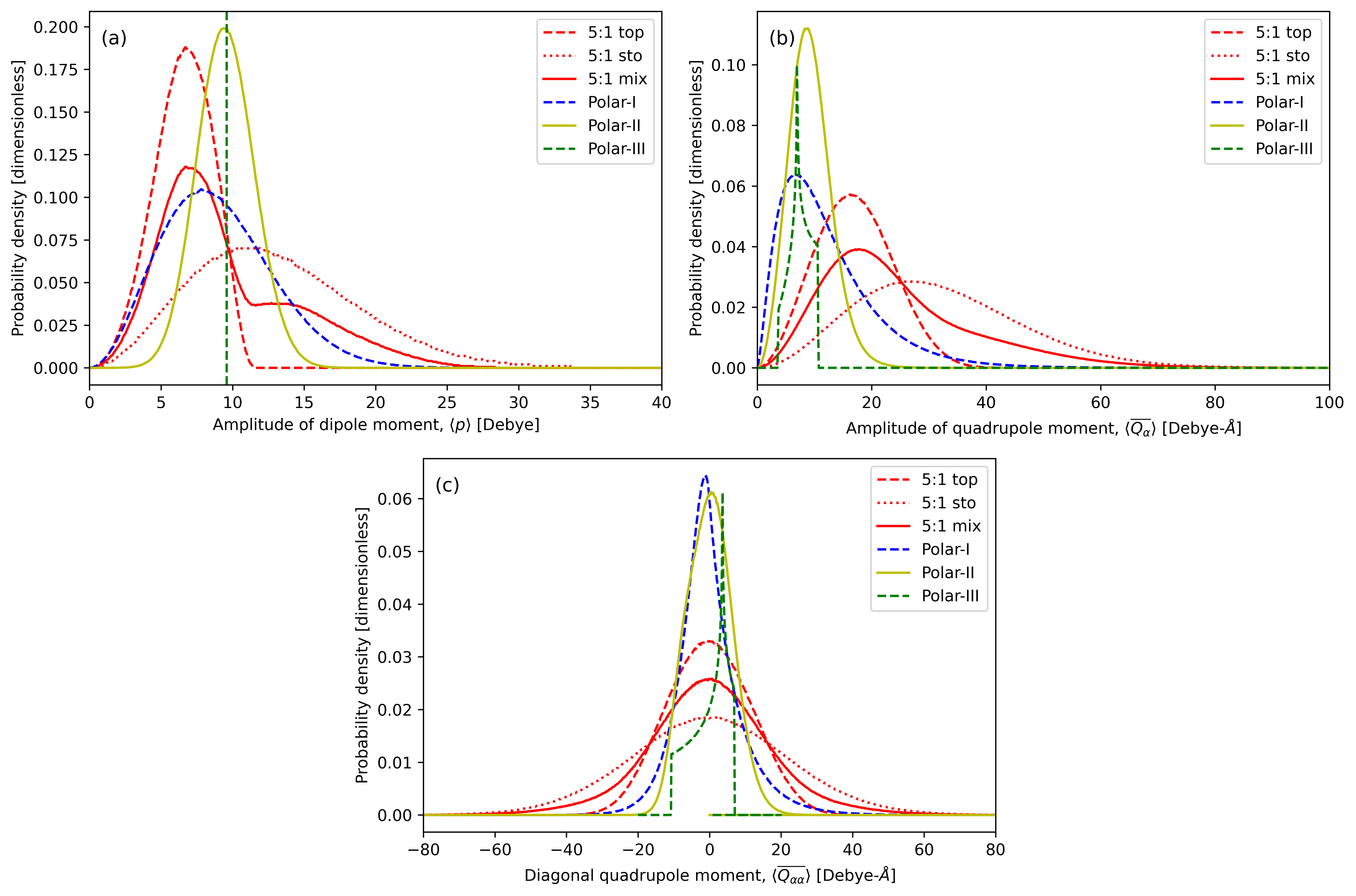}
    \caption{Probability distributions of (a) dipole moment magnitude, (b) average quadrupole moment amplitude and (c) average diagonal component of quadrupole moment (calculated in the global coordinate system), comparing between those derived by the three types of coarse-graining (molecular, stoichiometric and a 50/50 mixture) of TIP3P water simulations for clusters of 5 water molecules per bead with those obtained from simulations of the three proposed water models, \emph{Polar-I} (flexible), \emph{Polar-II} (constrained) and \emph{Polar-III} (rigid body).  The probability distribution values for the quadrupole moment distributions of the rigid body (\emph{Polar-III}) model have been rescaled to better fit in the plots alongside the other distributions.}
    \label{compare}
\end{figure}

\begin{figure}
    \centering
    \includegraphics[width=\columnwidth]{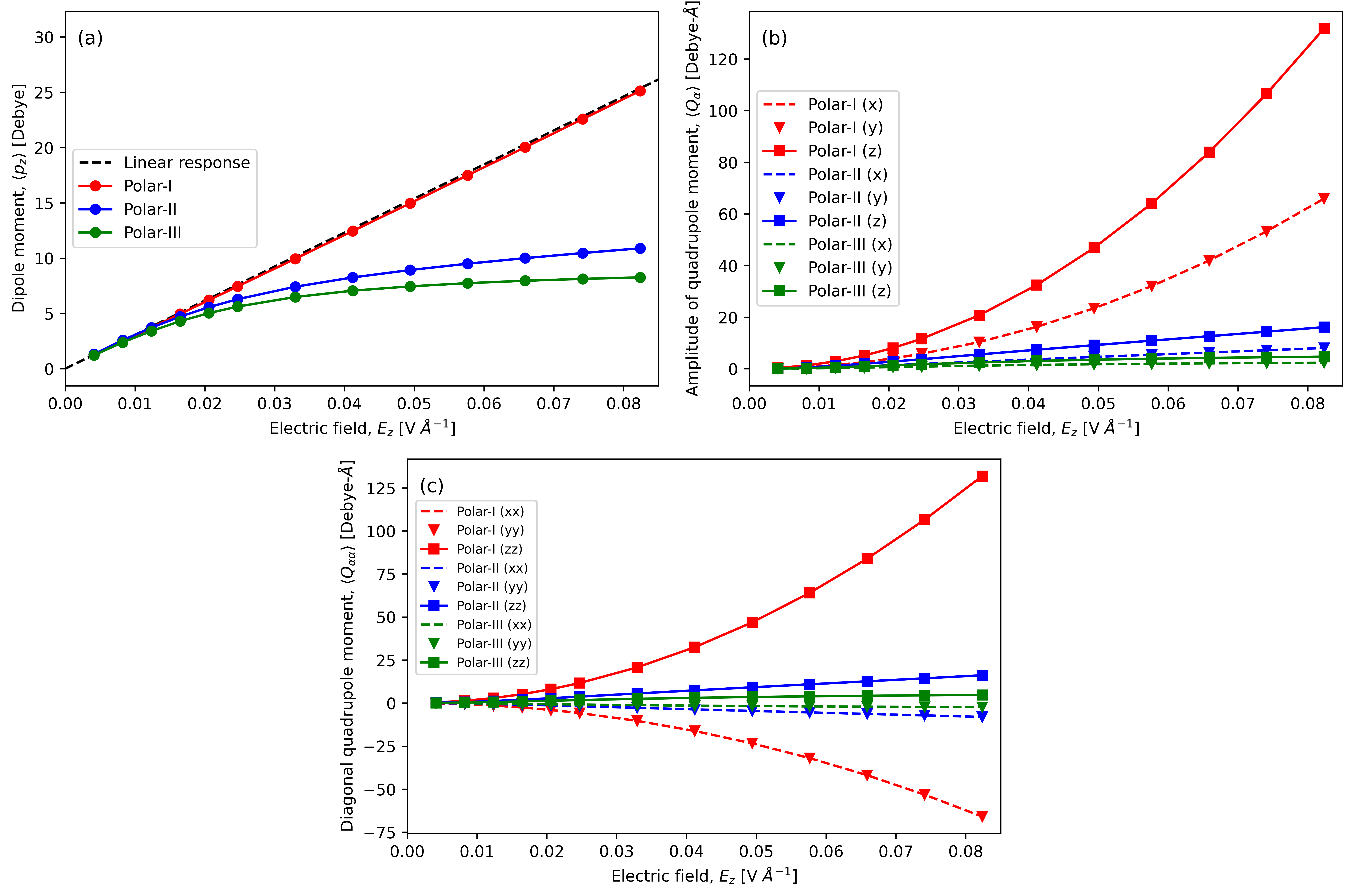}
    \caption{Induced polarisability response to external electric field of (a) dipole moment along field direction (compared with continuum electrostatic theory, dashed line), (b) quadrupole moment amplitudes and (c) diagonal components of quadrupole moment calculated in the global coordinate system, comparing between these obtained from simulations of the three proposed water models, \emph{Polar-I} (flexible), \emph{Polar-II} (constrained) and \emph{Polar-III} (rigid body).}
    \label{ext}
\end{figure}

Figure~\ref{compare} compares and contrasts the probability distributions of $\langle p \rangle$, $\langle\overline{Q_{\alpha}}\rangle$ and $\langle\overline{Q_{\alpha\alpha}}\rangle$ for the \emph{Polar-I} (flexible), \emph{Polar-II} (constrained) and \emph{Polar-III} (rigid body) models alongside the three 5:1 CG ones -- topological, stoichiometric and their 50/50 mix -- based on the TIP3P water model.  It is visually striking in Figure~\ref{compare}(a) that the $\langle p \rangle$ distribution of the Polar-I model matches that of the mixed CG closely in terms of general shape and expected average.  Their expected values $E(\langle p \rangle)$ are also very close, being within $1\%$ of each other. To provide context, the expected values for the different $\langle p \rangle$ distributions (all in units of Debye) are as follows: 6.26 (5:1 top), 13.62 (5:1 sto), 9.94 (5:1 mix), 8.96 (\emph{Polar-I}), 9.52 (\emph{Polar-II}) and 9.58 (\emph{Polar-III}).  We also note that $E(\langle p \rangle)$ emerging from our 4:1 mixed CG $\langle p \rangle$ distribution (9.5 Debye) exactly matches the values reported by Vaiwala \emph{et al.}\cite{vjt18} for their 4:1 DPD-based CG models of water.

Although we have not optimised our models to generate representative quadrupoles, Figure~\ref{compare}(b) shows an overlap of the $\langle\overline{Q_{\alpha}}\rangle$ distributions between the \emph{Polar-II} model and the topological CG view, both having a very symmetric distribution shape, which we believe is a consequence of the angular restraint in \emph{Polar-II}.  Again, it is the \emph{Polar-I} distribution that provides the best cover for the desired broadness, $Max(\langle\overline{Q_{\alpha}}\rangle)$, of the mixed 5:1 CG distribution.  The expected values $E(\langle\overline{Q_{\alpha}}\rangle)$ of the different models are as follows: 12.01 for \emph{Polar-I}, 4.64 for \emph{Polar-II} and 7.42 for \emph{Polar-III} as compared to 24.48 for the mixed 5:1 CG distribution, all in Debye$\cdot$\emph{\AA}.

Figure~\ref{compare}(c) shows a comparison of the $\langle\overline{Q_{\alpha\alpha}}\rangle$ distributions between the three polar models and the 5:1 CG methods of the TIP3P model.  Unsurprisingly, the broadness of all polar models' distributions decreases with decreasing the models' flexibility, and \emph{Polar-I} performs best at matching the broadness and shape of the mixed 5:1 CG of TIP3P distributions, whereas \emph{Polar-III} generates a single value for the $\langle p \rangle$ distribution and very narrow, numerically odd distributions for $\langle\overline{Q_{\alpha}}\rangle$ and $\langle\overline{Q_{\alpha\alpha}}\rangle$.

Finally, Figure~\ref{ext} represents our results of polarisability reactions and differences in performance for our three polar water models when subjected to an external electric field, $\vec{E}=\left(0,0,E_z\right)$.  It is clear from Figure~\ref{ext}(a) that the average induced dipole moment of the polar water bead along the direction of the external field, $\langle p_z \rangle$, increases with increasing $E_z$. We observe that this correlation starts off as a linear dependence of $\langle p_z \rangle$ on $E_z$ but deviates from this as the intensity of the electric field increases. The flexibility of the water model determines both the nature of this deviation from linear behaviour and the external field strength at which this occurs. For instance, \emph{Polar-I} produces the most persistent linear response for the entire range of $E_z$ values tested here, while \emph{Polar-II} and \emph{Polar-III} exhibit plateauing of induced dipole moments at respectively lower electric fields. These behaviours match up well to those observed for other CG polarisable water models: our most flexible \emph{Polar-I} model exhibits similarities to the `dressed solvent' model given in Chiacchiera \emph{et al.}\cite{chiacchiera2024}, while the more constrained \emph{Polar-II} model produces a response that strongly resembles the models devised by Vaiwala \emph{et al.}\cite{vjt18}. These similarities are striking, not least given the different background medium for the `dressed solvent' (oil instead of vacuum), the different CG degrees (3:1 for Chiacchiera \emph{et al.}, 4:1 for Vaiwala \emph{et al.}) and the use of 3CM in the selected models as opposed to 2CMs.

The linear response of our \emph{Polar-I} model compares well with the expected response of bulk liquid water according to classical continuum electrostatic theory, indicated as a dashed line in Figure~\ref{ext}(a):
\begin{equation}
\label{eq:pzez}
\langle p_z \rangle =  \frac{1}{\rho_m} \epsilon_0 \left( \epsilon_r - 1 \right) E_z = \alpha_p E_z,
\end{equation}
where $\rho_m$ is the number density of dielectric molecules (i.e. our polar water entities), and $\alpha_p \equiv \frac{1}{\rho_m} \epsilon_0 \left( \epsilon_r - 1 \right)$ is the polarisability for each water entity.  Use of this theory and the measured data points for this flexible water model predicts a relative permittivity for bulk water of $\epsilon_r \approx 76.7$, within the error for the value calculated using Eq.~\ref{dielec} ($78.1$).  This initial linear response compares well with Chiacchiera \emph{et al.}'s\cite{chiacchiera2024} `dressed solvent' model and confirms the fact that the charge flexibility of the model, and hence unhindered charge penetration, is crucial for achieving such a linear response.

\begin{figure}
    \centering
    \includegraphics[width=0.6\columnwidth]{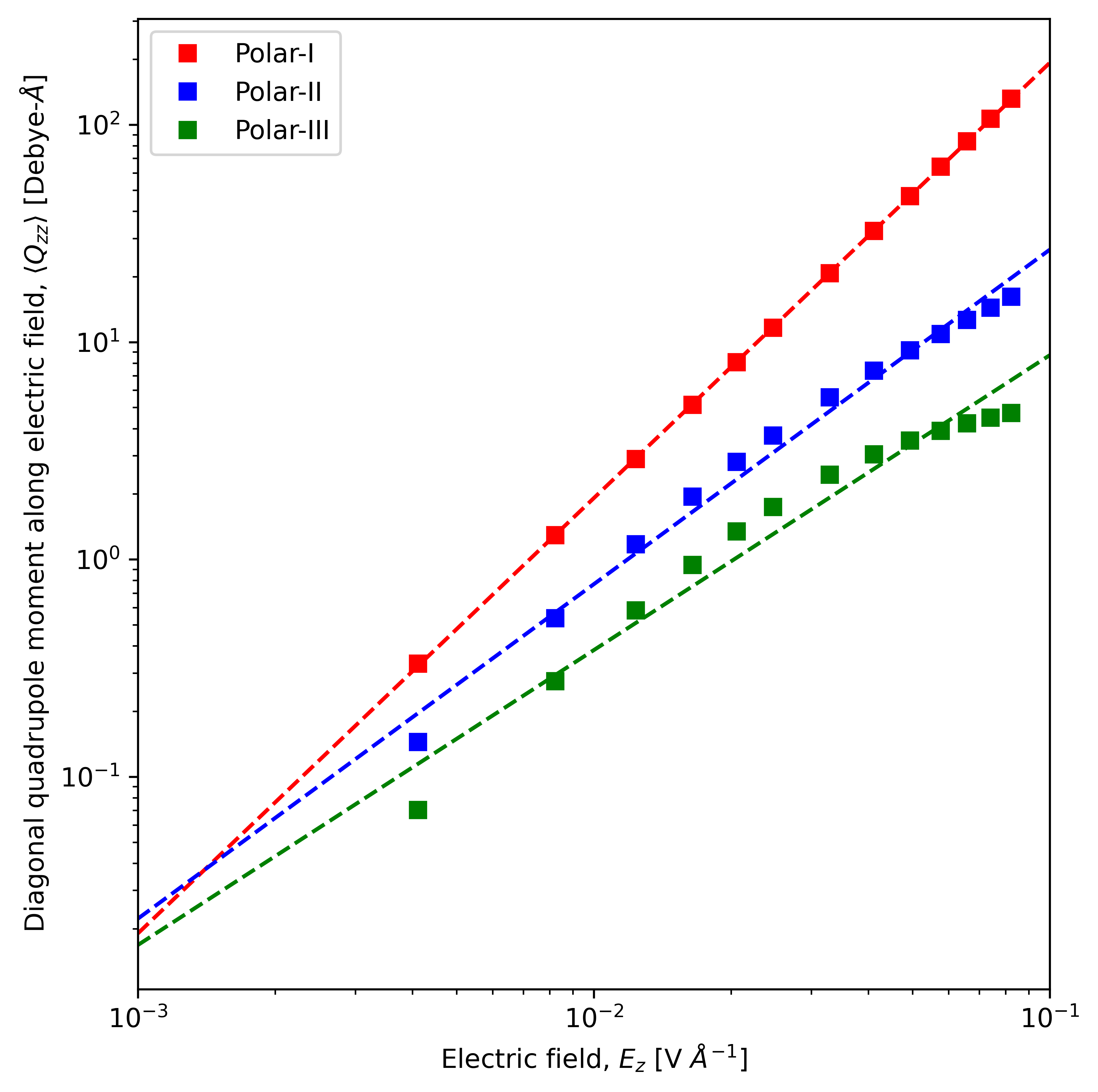}
    \caption{Logarithmic plots of the induced quadrupole moment response in the global coordinate system to an external electric field and in the direction of the field, obtained from simulations of the three proposed water models and comparing with their best fits to a power law equation (dashed lines): \emph{Polar-I} (flexible, $\langle Q_{zz}\rangle \approx 19372 E_z^{2.003}$), \emph{Polar-II} (constrained, $\langle Q_{zz}\rangle \approx 919.7 E_z^{1.539}$) and \emph{Polar-III} (rigid body, $\langle Q_{zz}\rangle \approx 197.7 E_z^{1.357}$).}
    \label{extlog}
\end{figure}

How the models behave in terms of their quadrupole tensor $\langle Q_{\alpha\beta} \rangle$ (Eq. \ref{qab}) when subjected to electric fields is also very interesting.  The averaged quadrupole moment per entity $\hat{Q}$, as calculated in the global frame of reference, represents the macroscopic nature of the dielectric medium and its constituent elemental `molecules'.  In the absence of an external electric field its diagonal elements are distributed around zero, as shown in Figure~\ref{compare}(c), with randomly and equally distributed amplitudes, \ref{compare}(b); this is as expected when the entities' internal charge assembly and their associated individual dipoles and quadrupoles are randomly oriented and distributed.  Upon application of a constant external electric field in the $z$-direction, $E_z$, the entities' response leads to a complete internal rearrangement of $\hat{Q}$, driven by reorientation of all of their internal dipoles to align with the external field, as it transforms into a diagonal form with a clear split between its components.  The quadrupole amplitudes for all three models split as $\langle Q_{z} \rangle~>~\langle Q_{x} \rangle~=~\langle Q_{y} \rangle$ -- as shown in Figure~\ref{ext}(b) -- with the amplitude in the direction of the electric field equal to the sum of the other two components, $\langle Q_{z} \rangle~=~\langle Q_{x} \rangle~+~\langle Q_{y} \rangle$.  The diagonal components of the quadrupole for all three models -- given in Figure~\ref{ext}(c) -- exhibit a similar split:  $\langle Q_{xx} \rangle~=~\langle Q_{yy} \rangle~<~-\langle Q_{zz} \rangle$, where  $\langle Q_{zz} \rangle~=~-\langle Q_{xx} \rangle~-~\langle Q_{yy} \rangle$, confirming the traceless nature of $\hat{Q}$ (Eq. \ref{qab}).  The diagonalisation of $\hat{Q}$ is easily discerned from Figures~\ref{ext}(b)~and~(c), showing clearly that $\langle Q_{\alpha} \rangle~=~\left| \langle Q_{\alpha\alpha} \rangle \right|$ for $\alpha~=~x,y,z$.  

Interestingly, we also find that the $zz$-components of $\hat{Q}$ scale with the external field at least approximately according to a power law, $\langle Q_{zz} \rangle = a E_{z}^k$, as indicated in Figure~\ref{extlog}. Among the three water models, the power law index $b$ seemingly decreases as the model becomes less flexible: for the given range of $E_z$, the most flexible model (\emph{Polar-I}) gives a quadratic response ($k\approx2$), while $k\approx 3/2$ for the more constrained one (\emph{Polar-II}) and $k\approx 4/3$ when rigid bodies are in use (\emph{Polar-III}). Further complexity in the quadrupolar response to external electric field is visible in Figure~\ref{extlog} for the more constrained water models, which indicate that the power law index $k$ decreases with increasing $E_z$; for instance, the rigid body \emph{Polar-III} model appears to respond quadratically for smaller electric fields but approaches linear behaviour ($k \approx 1$) for the largest values. In a similar manner to the induced dipole behaviour indicated by $\langle p_z \rangle$, the quadrupole response $\langle Q_{\alpha\beta} \rangle$ to the external field $E_z$ also diminishes in magnitude as the model's flexibility diminishes, as indicated by the values of $a$ for the fitted power law relationships. These trends very strongly confirm the need for flexibility in bonds attaching (Drude) satellite charges to enable them to probe charge distributions of their own water entities as well as those of any surrounding water `molecules', especially if the polarisation traits of water are to be represented efficiently and effectively.

As we discussed earlier when making the connection to molecular dynamics models via continuous coarse-graining, the $\hat{Q}$ response to a constant electric field is predominantly due to dipole-quadrupole coupling, $\alpha_Q\propto Q:Q/p.p$, emerging from localised electric field gradients generated by the induced dipoles and quadrupoles\cite{hohm2000}.  By comparing $E(\langle Q:Q/p.p \rangle)$  derived from the $\langle p \rangle$ and $\langle\hat{Q}\rangle$ distributions across the three polar water models -- 6.72, 1.58 and 1.90 for \emph{Polar-I, II} and \emph{III} respectively in units of \AA$^2$ -- with that from coarse-graining the TIP3P water model at a 5:1 CG level  -- 18.76 -- one can see that the \emph{Polar-I} model delivers the largest dipole-quadrupole coupling albeit smaller by a factor of 2.8 compared to that obtained from the 5:1 CG of the TIP3P model, whereas \emph{Polar-II} and \emph{Polar-III} generate much smaller ones.  We can re-interpret this comparison between the polar water models by examining the resulting entities' quadrupole lengths: defining the model's deviation as $\delta L_Q^{\textbf{model}}=L_Q^{\textbf{model}}/L_{Q,l_{\textbf{CG}}=5}-1$, where $L_{Q,l_{\textbf{CG}}=5}$ is the quadrupole length obtained from coarse-graining the TIP3P water model to the same CG lengthscale as our models ($l_{\textbf{CG}}=5$), we obtain deviations for \emph{Polar-I}, \emph{Polar-II} and \emph{Polar-III} of -40.2\%, -71.0\% and -68.2\% respectively.

The above observations strongly confirm that the flexibility of the 3CM assembly is critical for maximising the medium's ability to polarise.  Restrictions in bond flexibility either by imposing an equilibrium length and angle or full rigidification only lead to narrowing of the dipole and quadrupole distributions for the dielectric medium, thus limiting the model's capability to deliver the desired dielectric response both in the local environment and globally by increasing the decay of the linear response reaction of the dielectric medium to external electric fields.  The slightly improved average ability to polarise locally, $E(\langle Q:Q/p.p \rangle)$ and $E(\langle p\rangle)$, of the rigid model (\emph{Polar-III}) over that of the constrained model (\emph{Polar-II}) suggests that there may still be gains in exploring this space in future models. These could be achieved by aiming to preserve a more limited charge penetration across water entity assemblies via more careful modulation of the Drude bonds by, for instance, adding a quartic term to the harmonic potential.  All three models offer an internal quadrupole: that provided by \emph{Polar-I} is the most commensurate to that for a 5:1 CG representation of TIP3P.  This is a notable improvement on other advanced CG water models, which either lack this capability completely due to the adoption of a 2CM\cite{chiacchiera2024,riniker2011,vjt18} or compromise on polarisable performance, especially within the local environment, due to the adoption of a rigidified 3CM\cite{wu2010,li2020}.

\section{Discussion and Conclusions}

This study has generated polarisable coarse-grained water models that enable the bulk medium to behave as a dielectric \cite{chiacchiera2024}, so that liquid at the mesoscopic level can respond to changes in the local environment due to the presence of both ions and external electric fields while simultaneously preserving the thermodynamic and hydrodynamic behaviours of water \cite{seaton2024}.  In the spirit of mesoscopic modelling helping to bridge length and time scales between atomistic and continuum levels, the DPD models had to both (i) be minimalistic in terms of force field parameters to reduce computational requirements on the one hand, and on the other (ii) preserve the polarisability traits of already established atomistic water models, such as TIP3P, as much as possible. 

The latter requirement led to a new discovery while exploring different strategies for coarse-graining (CG) data from atomistic trajectories.  The dualistic nature of the mesoscopic scale emerged from a simple atomistic model, enabling two distinct representations -- topological (molecular) and (volumetrically minimal) stoichiometric -- of polarisablity distributions for the same CG level.  While this finding imposes a requirement on mesoscopic models to be inclusive of both distributions, it also helps us to compare, contrast and, ultimately, judge the goodness of representations between different mesoscopic models.  The information from the consistent level-by-level coarse-graining of our `ground truth' atomistic model (TIP3P) provides crucial evidence of the inevitable loss of information -- locality of structure and entropy -- due to aggregating molecules within beads and integrating out the degrees of freedom, despite retaining all original \emph{atomistically accurate} charge distributions in the background which determine the dielectric properties of medium represented by the model.  The effects of that loss for each CG level are quantified as built-in errors (relative compromises for the CG representation) to key polarisability-related properties, given in Table~\ref{tab:cgl}, which impose unsurpassable, inherently intrinsic limits on the inclusion of polarisation into any mesoscopic model when coarse-graining the electronic and/or atomistic degrees of freedom.  These errors also indicate a particular inefficiency in the inclusion and representation of polarisability for water models at lower CG levels (below 5:1) and the trend of diminishing returns when attempting to improve upon this error at higher levels (9:1 and beyond).

Figures \ref{add}, \ref{aqd} and \ref{aqdi} clearly demonstrate an emergent dualistic nature for the distributions of the dipole moment magnitude ($\langle p \rangle$), the average quadrupole amplitude ($\langle\overline{Q_{\alpha}}\rangle$) and averaged diagonal components ($\langle\overline{Q_{\alpha\alpha}}\rangle$) of the quadrupole moment (as calculated in the global coordinate system).  The common observation in these is that for the same coarse-graining level, the distribution spread of the stoichiometric representation is approximately twice as broad as that of the topological one, which we have attributed to more favourable inclusion of hydrogen bonding within the stoichiometric representation.  This same observation with respect to distribution wideness is also valid for the expected distributions of dipole moment magnitudes and the quadrupole moment average amplitudes.  We averaged the distribution data from the two distinct CG approaches as a 50/50 mix, representing each equally from a mesoscopic point of view.  This has resulted in new mixed distributions that each inherit the broadness of the stoichiometric representation and produce a peak closer to that obtained from the topological representation.  We also demonstrated that within the continuous series of CG levels examined -- 2:1 to 13:1 -- the 5:1 level would be a very good choice for a CG model of water since it generates the minimum mismatch between the two distinct CG representations for dipole distributions and quadrupolarisability.  It is also a very sensible choice in terms of the water bead representing and emulating solvation shells around ions, specifically those found in vanadium redox flow battery electrolytes.  Also, as exemplified in Table~\ref{tab:cgl}, any further CG beyond 7:1 level is unlikely to bring any further benefits in minimising the overlap of essential dielectric properties distributions between the two types of coarse-graining we described above.   

Based on the requirements and the new knowledge outlined above, we devised three different minimal CG models for water (parameterised in Table~\ref{tab:params}). Starting with a three-charge model (3CM) with flexible bonds between central beads and Drude charges (\emph{Polar-I}), we subsequently constrained the inter-bead assembly with stiffer bonds and an angle potential to provide one alternative (\emph{Polar-II}), before rigidifying it to provide another (\emph{Polar-III}). In all three models, we preserved the general force field interaction type that we previously used for our non-polar water model -- \nDPD\ using $n=4$ -- but conveniently adjusted its energy scale to recover at least some of its macroscopic thermodynamic and hydrodynamic properties. By doing so, all three CG polar water models predict the correct liquid density at room temperature and commensurately similar vapour densities compared with the non-polar model, as well as obtain identical self-diffusivities and shear viscosities, both of which are additionally adjustable via the DPD pairwise thermostat. A noticeable decrease in interfacial tension between phases was observed when the charges are added, but there is still scope to adjust the \nDPD\ interactions further to correct for this and the overly high vapour density. 

We tested the fitness of our three models to represent the polarisability traits of an established atomistic water model, TIP3P, by comparing and contrasting the models' polarisability distributions against those that emerge when coarse-graining an atomistic trajectory to a level of 5:1.  As demonstrated in Figure~\ref{compare}, our best model -- \emph{Polar-I} -- produces a dipole moment magnitude distribution that matches very well against the desired mixed 5:1 coarse-grained trajectory of the TIP3P atomistic model.

Given the choice of polarisable model (3CM) and unit test charge, we recover distributions of ensemble-averaged quadrupole moment per entity that, in terms of both magnitudes $\langle\overline{Q_{\alpha}}\rangle$ and diagonal tensor components $\langle\overline{Q_{\alpha\alpha}}\rangle$, match reasonably well with those obtained from coarse-graining the TIP3P water model at a 5:1 level.  In particular, the \emph{Polar-I} model produces s quadrupole moment distribution that closely matches the shape obtained from the mixed 5:1 CG representation of TIP3P, while \emph{Polar-II} reasonably closely matches the shape of the topological CG representation.  Our rigid model (\emph{Polar-III}) performs as intended, still offering polarisation to the system via a fixed dipole and narrowly-distributed quadrupole moments.

%The virials $W$ of different types of particle interactions a model system provide information of the relationship between energy and structure in it.  Therefore, we could gage the extent of influence of \nDPD of interaction on the dielectric properties of the material to explore the room for improvement of our model in the future.  Focusing only on our \emph{Polar-I} model the virials of the \nDPD interaction ($W_i$), electrostatics ($W_e$) and Drude harmonic bonds ($W_b$) interactions are approximately $W_i=W_e=W_b/3$ as while the total systems energy of the system was dominated by two orders and It is worth noting that we tested virtually what is the effect on $\langle p \rangle$ and $\langle\overline{Q_{\alpha}}\rangle$ if $A_ij$ was increased 5-fold or reduced to 0 (i.e. no bead-bead interactions, and the electrostatics dominates)

% Share the size of Eng and Vir of nDPD vs elc vs bonding and the effect of removing nDPD or magnifying nDPD's A lead to further polarisation for the same artificial substance.

The goodness of the three models was ultimately tested by the application of an external electric field, $E_z$. Figure~\ref{ext} demonstrates that the models' induced polarisability reaction increases with increasing $E_z$, while it diminishes when a model's flexibility is decreased. This confirms the superiority of \emph{Polar-I} over its constrained and rigidified versions, as well as the correctness of our chosen approach to include polarisation into our original non-polar water model.  The polarisability reaction of \emph{Polar-I} to external electric fields is an excellent fit to expected behaviour, given that the CG water entity's induced dipole responds linearly to the electric field and nearly perfectly according to the predicted response given by classical continuum electrostatic theory (Eq. \ref{eq:pzez}): this level of agreement is also attained by the underlying atomistic models\cite{cox2020}. Our \emph{Polar-I} model's induced dipole response also retains linear behaviour for larger external electric field intensities, which is neither the case for our \emph{Polar-II} and \emph{Polar-III} models nor for other polar models\cite{vjt18}.  Furthermore, our models, in particular \emph{Polar-I}, also generate \emph{by construction} an excellent, non-linearly induced quadrupole response to external electric fields that increases with the field intensity. It is this quadrupole response of molecular water that, as substantiated by Gongadze \emph{et al.}\cite{gongadze2013}, facilitates re-arrangement of water molecules and their relative re-orientation with respect to an applied external field.  

We note that the entity's (molecular) quadrupole in a polar solvent provides a critical interaction with the surrounding medium --  especially near interfaces involving charges, phase coexistence, bio-membranes, etc. -- leading to concerted local quadrupole distributions\cite{gongadze2013} that are crucial for modulated transmission of electrostatic interactions across the interfaces.  By lowering the dielectric permittivity locally near such interfaces, these localised quadrupole moments ultimately affect emergent interface properties such as the equilibrium densities of coexisting vapour and liquid phases, surface tension and shear viscosity.  We note that the dipole-quadrupole coupling of our \emph{Polar-I} model is the closest commensurately to that of the TIP3P water model at a CG level of 5:1, while the couplings produced by \emph{Polar-II} and \emph{Polar-III} are seriously under-emphasised, thereby confirming that future improvements should focus on flexible models.

\ack{The authors would like to thank our colleagues Silvia Chiacchiera, Vlad Sokhan and Patrick Warren (UKRI STFC Daresbury Laboratory) for useful discussions and our apprentice Tristan Rimmer for carrying out vapour-liquid equilibria and diffusivity simulations.}

\funding{This work was supported by the BatCAT consortium, funded under the UK government’s Horizon Europe funding guarantee -- grant number 10091190 -- and by European Union’s Horizon Europe research and innovation programme -- grant agreement number 101137725 -- and made use of computing resources on ARCHER2 provided by CoSeC, the Computational Science Centre for Research Communities, and SCARF by STFC Scientific Computing.}
% This section is a list of funder names and grant numbers

\roles{\textbf{Michael A. Seaton}: Conceptualization, Methodology, Software, Validation, Formal analysis, Investigation, Writing (Original Draft), Visualization, Supervision; \textbf{Benjamin T. Speake}: Software, Validation, Formal analysis, Investigation, Writing (Review \& Editing), Visualization; \textbf{Ilian T. Todorov}: Conceptualization, Methodology, Validation, Resources, Writing (Review \& Editing), Supervision, Project administration, Funding acquisition.}
% List author names and the contributions made to the article, using terms from the NISO Contributor Roles Taxonomy (CRediT) https://credit.niso.org

%\data{Sample text inserted for demonstration.}
% For more information on IOP Publishing's research data policy see: https://publishingsupport.iopscience.iop.org/questions/research-data/

%\suppdata{Sample text inserted for demonstration.}

%\section*{References}
\bibliographystyle{iopart-num}
\bibliography{references}% Produces the bibliography via BibTeX.

\end{document}